\documentclass{aa}  

\usepackage{graphicx}
\usepackage{txfonts}
\begin{document} 

\title{The GTC exoplanet transit spectroscopy survey I}
\subtitle{OSIRIS transmission spectroscopy of the short period planet WASP-43b}
 
  \author{F. Murgas
          \inst{1}\fnmsep\inst{2}
          \and
          E. Pall\'{e}\inst{1}\fnmsep\inst{2}
          \and
          M. R. Zapatero Osorio\inst{3}
          \and
          L. Nortmann\inst{4}
          \and
          S. Hoyer\inst{1}\fnmsep\inst{2}
          \and
          A. Cabrera-Lavers\inst{1}\fnmsep\inst{2}
          }

   \institute{Instituto de Astrof\'isica de Canarias (IAC), E-38205 La Laguna, Tenerife, Spain\\
              \email{murgas\_ext@iac.es}
         \and
             Departamento de Astrof\'isica, Universidad de La Laguna (ULL), E-38206 La Laguna, Tenerife, Spain
         \and
             Centro de Astrobiolog\'ia (CSIC-INTA), E-28850 Torrej\'on de Ardoz, Madrid, Spain
         \and 
             Institut f\"ur Astrophysik, Georg-August-Universit\"at, Friederich-Hund-Platz 1, D-37077 G\"ottingen, Germany
             }

   \date{Received July 2013; Accepted January, 2014}

 
  \abstract
   {Of the several extrasolar planets discovered to date, only a few of them have orbital periods of less than a day. 
   Such planets are interesting candidates to study tidal effects and, in the case of short period Hot Jupiters, 
   they offer an excellent opportunity to detect and study their atmosphere due to their generally large atmospheric scale heights.}
   {In this work, we use long-slit spectroscopy observations of a transit event of the close-in orbiting planet WASP-43b 
   ($M_p = 2.034$ $M_{Jup}$, $R_p = 1.036$ $R_{Jup}$) in an effort to detect its atmosphere.}
   {We used Gran Telescopio Canarias (GTC) instrument OSIRIS to obtain long-slit spectra in the optical range 
   (520-1040 nm) of the planetary host star WASP-43 (and a reference star) during a full primary transit event and four 
   partial transit observations. We integrated the stellar flux of both stars in different wavelength regions producing several 
   light curves. We fitted transit models to these curves to measure the star-to-planet radius ratio, $R_p/R_s$, across 
   wavelength among other physical parameters.}
   {We measure a mean planet-to-star radius ratio in the white light curve of $0.15988^{+0.00133}_{-0.00145}$. 
   Using broad band filters, we detect the color signature of WASP-43.
   We present a tentative detection of an excess in the planet-to-star radius ratio around the Na\,{\sc i} doublet ($\lambda$ 588.9, 589.5 nm) when compared to the nearby continuum at the 2.9-$\sigma$ level.
   We find no significant excess of the measured planet-to-star radius ratio around the K\,{\sc i} doublet ($\lambda$ 766.5 nm, 769.9 nm) 
   when compared to the nearby continuum. Combining our observations with previous published epochs, we refine the estimation of the 
   orbital period. Using a linear ephemeris, we obtained a period of $P=0.81347385 \pm 1.5 \times 10^{-7}$ days. 
   Using a quadratic ephemeris, we obtained an orbital period of $0.81347688 \pm 8.6 \times 10^{-7}$ days, and a change in this parameter of $\dot{P} = -0.15 \pm 0.06$ sec/year. As previous results, this hints to the orbital decay of this planet although a timing analysis over several years needs to be made in order to confirm this.
   }
   {}


   \maketitle
%

\section{Introduction}

Among the increasing family of extra-solar planets discovered so far, only a handful of them 
possess orbital periods of less than a day. Some examples of close-in exoplanets includes possible disintegrating 
planets (e.g. KIC 12557548b, \citealp{Rappaport2012}), super-Earths (e.g. Corot-7b, \citealp{Leger2009}; 55 Cnc e 
\citealp{McArthur2004}, \citealp{Dawson2010}, \citealp{Winn2011}), and hot Jupiters (e.g. WASP-18b \citealp{Hellier2009}; WASP-19b \citealp{Hebb2009}; and WASP-43b \citet{Hellier2011}). The study of such close-in orbiting planets 
can shed some light in the dynamic interactions that produce planetary migrations, tidal interactions between the star and the 
planet, and in the case of transiting hot Jupiters, they offer a good opportunity to study the atmospheric composition of 
extra-solar planets under extreme stellar irradiation.

WASP-43b was discovered by the Wide Angle Search of Planets team in 2011 (\citealp{Hellier2011}). It orbits a K7V star 
with an estimated mass of $0.717 \pm 0.025 M_\odot$ and a stellar radius of $0.667 \pm 0.011 R_\odot$ 
(\citealp{Gillon2012}). \citet{Gillon2012} presented a study of several transits observed with TRAPPIST, VLT near-IR 
photometry, and CORALIE radial velocities. They improved the parameters of the system finding an eccentricity 
consistent with a circular orbit ($e=0$), a semi-major axis over stellar radius of $a/R_s = 4.918^{+0.053}_{-0.051}$, 
and an estimated planetary mass and radius of $2.034 \pm 0.052 M_{Jup}$ and $1.036 \pm 0.019 R_{Jup}$, respectively. 
By using secondary eclipse observations, the thermal emission of the planet at 1.16 $\mu m$ and 2.09 $\mu m$ was 
detected. \citet{Blecic2013} measured the secondary eclipse of WASP-43b in the 3.6 $\mu m$ and 4.5 $\mu m$ bands 
using Spitzer, deducing a planetary brightness temperature of $1684 \pm 24$ K and $1485 \pm 24$ K respectively. 
\citet{Wang2013} also detected the secondary eclipse in $H$ and $K_s$ bands from the ground using the Wide-field
Infrared Array Camera (WIRCam) mounted at the Canada-France-Hawaii Telescope (CFHT); their data are consistent 
with a blackbody with a temperature of 1850 K, higher than the expected equilibrium temperature of the planet. 

\citet{Blecic2013} also improved the estimation of the orbital period by a factor of three compared to \citet{Gillon2012}, 
finding a period of $P = 0.81347459 \pm 2.1 \times 10^{-7}$ days. With such a short period of less than a day, 
WASP-43b is expected to move towards smaller orbits over time, and eventually fall into the star in a time scale 
between 8-800 Myr depending on the efficiency of orbital energy dissipation from the star (\citealp{Hellier2011}). 
Due to its close orbit around its host star and its considerable size when compared to the stellar radius, 
WASP-43b offers a good opportunity to study this planet's atmosphere using current ground based instrumentation.

One of the most successful techniques to infer the atmospheric composition of extra-solar planets is 
transmission spectroscopy. When a planet transits its host star, part of the starlight will be absorbed 
by the atoms and molecules present in the planetary atmosphere. Since atoms and molecules absorb radiation 
at specific energies, they will produce variations in the flux transmitted through the atmosphere. 
A direct consequence of such flux variations will be the change in the measured depth of the transit at specific wavelengths. 
Hot jupiters are excellent targets for transmission spectroscopy studies, mainly because given their relative big sizes compared 
to their host star they produce a deep transit and, since they are gaseous planets, they possess big atmospheres with considerable 
scales heights. Some successful examples of transmission spectroscopy has been 
the detection of Sodium from space and ground in the systems HD 209458b (\citealp{Charbonneau2002}, \citealp{Snellen2008}) and 
HD 189733b (\citealp{Redfield2008}); the study of short period hot jupiters in the near-IR (\citet{Bean2013}), and 
the study of super-earths (e.g. \citealp{Bean2010}, \citealp{Bean2011}). 

Here, we probe the atmosphere of WASP-43b by means of transmission spectroscopy of one full and four partial planetary transit 
events. In \S \ref{sec:obs} we describe the observation setup; in \S \ref{sec:meths} 
we describe the data reduction process to produce the final light curves, and the light curve fitting methodology; 
in \S \ref{sec:results} we present the results of the data analysis; and in \S \ref{sec:conclusions} the conclusions of 
our work.


\section{Observations}
\label{sec:obs}

The Gran Telescopio Canarias (GTC) is a 10.4 m telescope located at Observatorio Roque de los Muchachos 
in La Palma. GTC instrument OSIRIS (Optical System for Imaging and low Resolution Integrated Spectroscopy, \citealp{Cepa2000}) 
consists of two CCD detectors with a field of view (FOV) of 7.8 $\times$ 7.8 arcmin and plate scale of 0.127 arcsec/pix. GTC/OSIRIS 
has already been used successfully for planetary science in the past, using broad band photometry (e.g. \citealp{Tingley2011}), tunable 
filters observations (e.g. \citealp{Sing2011}, \citealp{Colon2012a}, \citealp{Murgas2012}), and long-slit transmission spectroscopy 
(e.g. \citealp{Sing2012}).

For our observations, we chose the $2 \times 2$ binning mode, a readout speed of 200 kHz with a gain of 0.95 e-/ADU and 
a readout noise of 4.5 e-. We used OSIRIS in its long-slit spectroscopic mode, selecting the grism R1000R which covers 
the spectral range of 520-1040 nm, and a custom built slit of 12 arcsec of width. With such a wide slit, the spectral 
resolution is dominated by the seeing during the observations, which varied between 0.8 and 1.8 arcsec. This translates in 
an effective spectral resolution of $R=$ 374-841 at 751 nm. The use of a wider slit has the advantage of reducing the possible 
systematic effects that can be introduced by light losses due to changes in seeing, airmass, and/or losses caused by differential 
telescope guiding between the two stars.

Data for WASP-43b were taken in visiting mode on January 8, 2013. As reference we chose 2MASS J10200126-0948099 ($J=10.16$ mag, $K_S=9.62$ mag), a star with a similar R band magnitude and located at a distance of 5.74 arcmin from the target. Due to the FOV of OSIRIS, 
the target and reference star had to be located in separate CCDs. The position angle 
of the reference star with respect to the target (measured east of north) was $PA = 2.39^\circ$. Observations began at 02:28 UT 
(one hour before ingress) and ended at 05:30 UT (one hour after egress). The exposure time was set to 15 seconds and the readout 
time of the instrument was close to 21 seconds, meaning we collected approximately one spectrum every 36 seconds. Spectra of both 
target and reference star were not collected at parallactic angle. However, given the large size of the slit width, the red wavelengths 
covered by our instrumental setup, and the airmasses of the observations ($\sec z=$ 1.45-1.35 with a minimum at 1.28 near 
mid-transit), no significant flux losses due to the Earth's atmospheric refraction are expected. Nevertheless, because the 
seeing varied during the observations, we applied a correction as explained in section \ref{sec:meths}.

Besides the data taken in January 2013 (set 5 from now on), we have four more transit observations taken with GTC/OSIRIS in 
December 22 2011 (set 1), February 8 2012 (set 2), February 21 2012 (set 3) and April 18 2012 (set 4). 
These observations were made using a different electronic configuration: the readout speed was set to 500 kHz 
with a gain of 1.46 e-/ADU and a readout noise of 8.0 e-. We decided to use this readout configuration to obtain a higher time cadence. Unfortunately, none of these observations were of the expected 
quality due to several issues. In particular, during the observations of set 1, 3, and 4 part of the flux was lost or the 
observations were interrupted due to technical problems with the telescope, while set 2 was observed under bad weather 
conditions. These old data sets were not adequate to perform a transmission spectroscopy study, but they can be used 
to search for the color signature variation from transit to transit, and to fit a transit model to the 
white light curves in order to obtain the central time of the eclipse. For posterior observations we changed to the 200 kHz readout speed to minimize the electronic noise associated with a faster read while not sacrificing too much the cadence of the data acquisition. This read mode has become the standard for GTC long-slit spectroscopy observations.


\section{Methods}
\label{sec:meths}
\subsection{Data reduction}
The raw data reduction was made using standard procedures. We combined bias 
and flat images using the Image Reduction and Analysis Facility (IRAF\footnote{IRAF is distributed by the 
National Optical Astronomy Observatory, which is operated by the Association of Universities for Research in 
Astronomy (AURA) under cooperative agreement with the National Science Foundation.}) in order to produce the 
basic calibration frames, and used these calibration images to correct the science frames before the extraction 
of the spectra.

The extraction and wavelength calibration was made using a PyRAF\footnote{Python environment for IRAF} script 
written for GTC/OSIRIS long-slit data. This script automatized some of the steps to produced the spectra such as: 
extraction of each spectrum, extraction of the corresponding calibration arc, wavelength calibration (using the HgAr, 
Xe, Ne lamps provided for the observations) and Doppler shift to take into account the possible shifts in wavelength 
due to the different radial velocities of the target and the chosen reference star. Given the low spectral resolution 
of our data, small to moderate errors in wavelength calibration does not introduce large uncertainties in the results 
presented here.

   \begin{figure}
   \centering
   \includegraphics[width=\hsize]{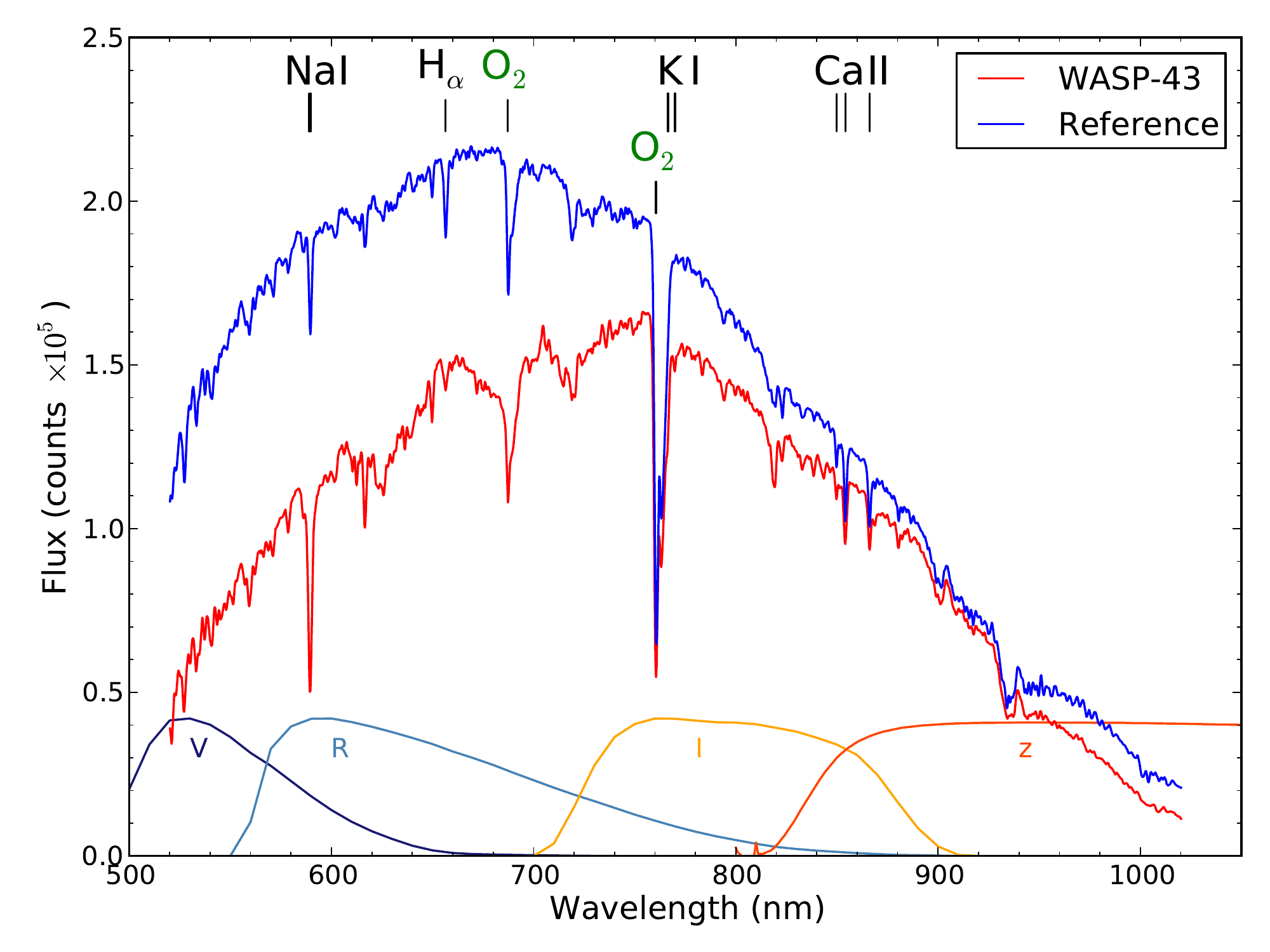}
      \caption{Extracted R1000R grism spectrum of WASP-43 (red) and its reference star (blue). The spectra 
      are not corrected for instrumental response or flux calibrated. Some stellar (black) and atmospheric O$_2$ telluric features 
      are shown (green). The V, R, I and \textit{z} sloan filters are also plotted in an arbitrary scale to show the wavelength 
      coverage of the spectrum.}
         \label{FigExtrSpec}
   \end{figure}

Several apertures for the spectra optimal extraction were tested during the reduction process, the one that delivered the 
best results in terms of low scatter measured using the root mean square (RMS) of the points outside of transit (after normalization) for the white light curve was selected. The results presented here for data set 5 were obtained using an aperture of 50 pixels of width, which corresponds to 12.7 arcsec on the detector. This is 7 to 14 times the raw seeing during the observations. 

The UT time of data acquisition was obtained using the recorded headers of the spectra 
indicating the opening and closing time of the shutter in order to compute the time of mid exposure. Then, we used 
the code written by \cite{Eastman2010}\footnote{http://astroutils.astronomy.ohio-state.edu/time/} to compute 
the Barycentric Julian Date in Barycentric Dynamical Time (BJD) using the mid exposure time for each of the spectra 
to produce the light curves analyzed here. 

Figure \ref{FigExtrSpec} presents an example of an extracted spectrum of WASP-43 and the reference star used in 
the time series. Both spectra are neither corrected for instrumental response nor flux calibrated.


\subsection{Filter definition}
The light curves were obtained using the numerical method known as Simpson's rule (\citealp{Numerical2002}) to integrate 
the fluxes of each spectrum of the time series and take the ratio between the integrated flux of the target and the 
reference star. To explore the presence of atmospheric features in WASP-43b, we created several filters of 
different width in wavelength domain.

\begin{figure}
   \centering
   \includegraphics[width=\hsize]{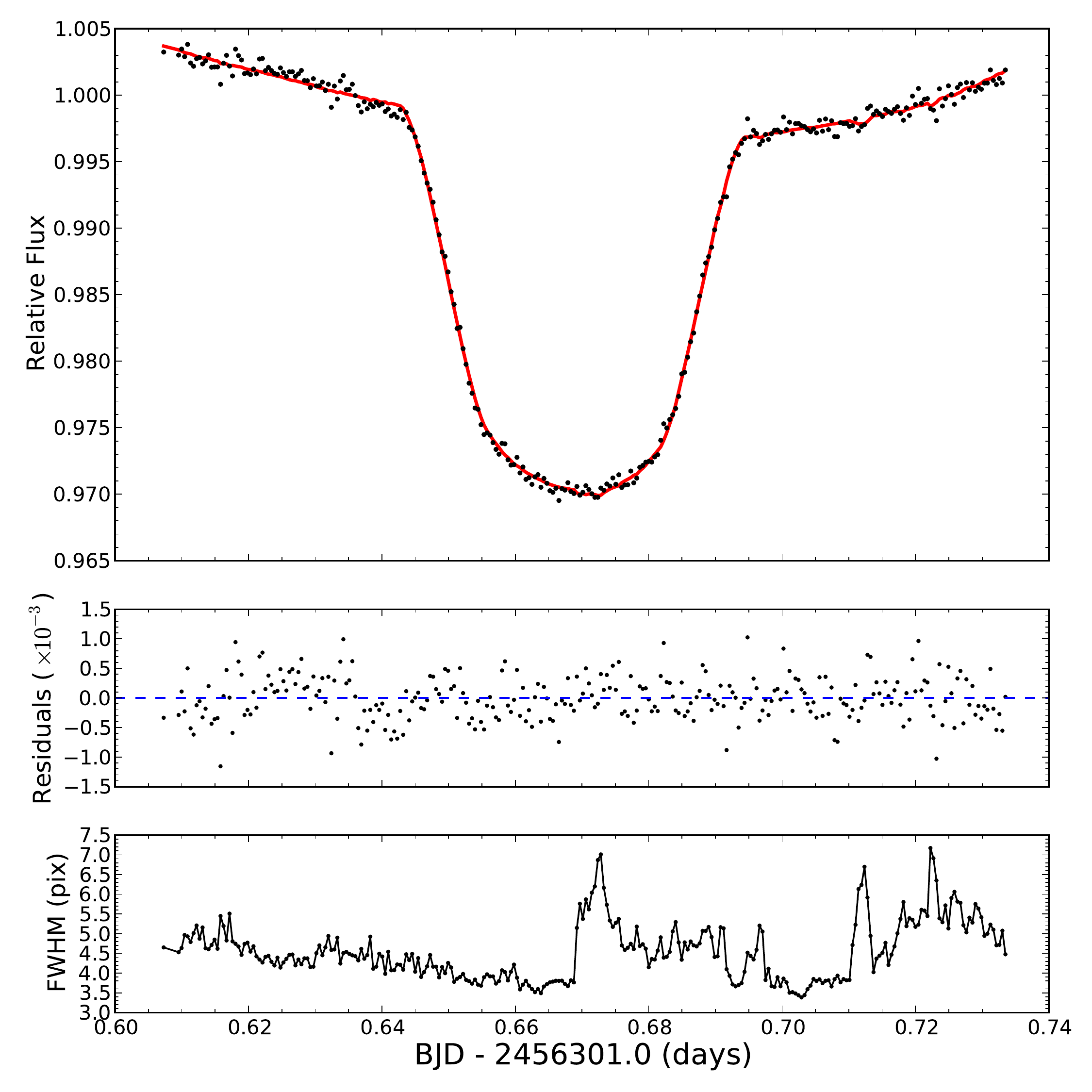}
      \caption{\textit{Top panel}: GTC/OSIRIS WASP-43b transit white light curve. 
      The red lines represents the best fit determined using our MCMC analysis. \textit{Middle panel}: 
      residuals of the fit. \textit{Bottom panel}: Full width at half maximum (FWHM) of WASP-43 in the spatial 
      direction versus time, showing the seeing variations during the observations.}
         \label{FigWLCurve}
   \end{figure}

\begin{figure}
   \centering
   \includegraphics[width=\hsize]{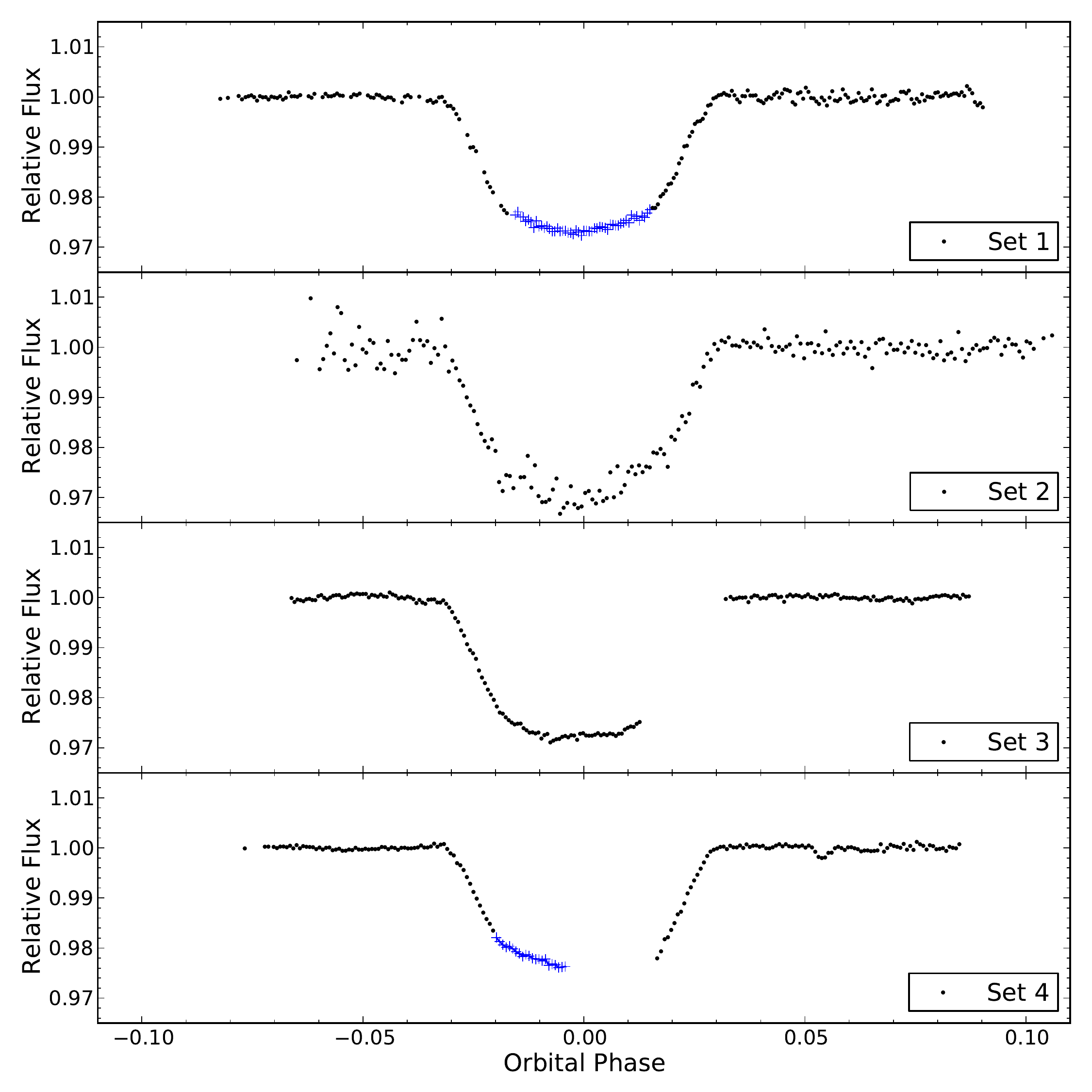}
      \caption{GTC/OSIRIS white light curves for WASP-43b for data sets 1 to 4. Blue crosses in sets 1 and 4 
      show data which were taken when the telescope presented technical issues.}
         \label{FigWLCAll}
   \end{figure}

In the case of the white light curve, the flux was integrated between 530 and 900 nm. Figure \ref{FigWLCurve} presents 
the white light curve for the data set 5 and Fig. \ref{FigWLCAll} shows the white light curves for the previous 
GTC/OSIRIS sets.

\begin{figure*}
   \centering
   \includegraphics[width=\hsize]{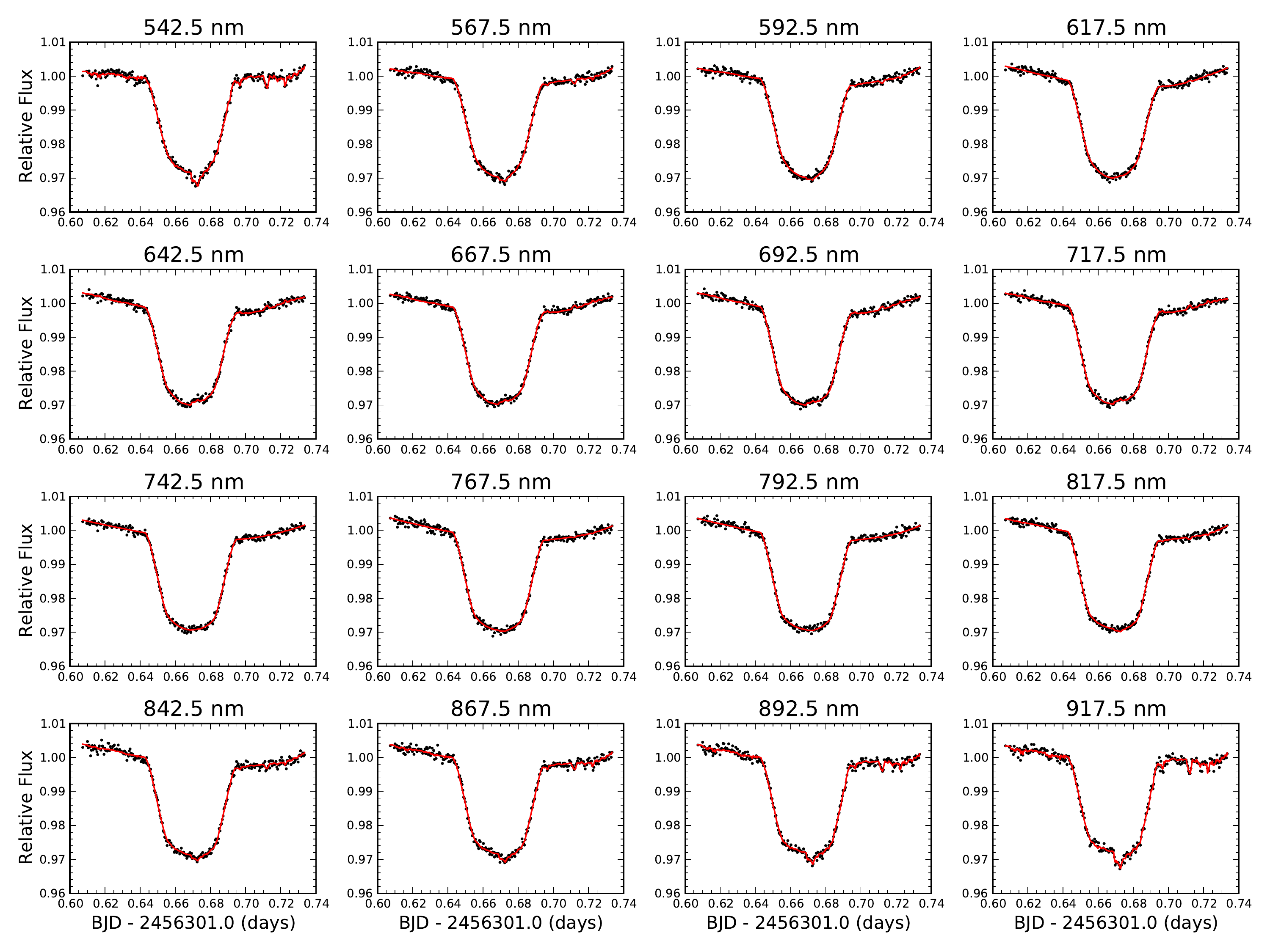}
      \caption{Light curves obtained using the filters of 25 nm of width for GTC set 5. 
      The red line shows the best fit found by the MCMC analysis.}
         \label{FigFilters}
   \end{figure*}

In order to get the transmission spectrum of WASP-43b, the observed spectra of WASP-43 and its reference star were 
integrated using filters of 25 nm of width in a wavelength range between 530 nm and 900 nm, resulting in 16 
light curves to be fitted (see Fig. \ref{FigFilters}). To test the robustness of the results found with the filters of 25 nm of 
width, we produced 34 light curves using filters of 10 nm of width with a wavelength range of 549.5-882 nm. The 10 nm bin centered at 589.6 nm has a width big enough to cover the Na\,{\sc i} 588.9 and 589.5 nm doublet. Besides these 25 and 10 nm bins, we produced 9 curves using fitlers of 18 nm width to cover the region near the K\,{\sc i} 766.5 and 769.9 nm doublet (wavelength range 702.0-855.5 nm). 

To study the color signature of WASP-43, we defined two broad filters one in the blue (530-680 nm) and in the red 
end of the observed spectra (800-950 nm). The filter coverage for the color signature is illustrated in the top panel of Figure 
\ref{FigColSign}. 

\subsection{Light curve fitting}
In order to get the transit parameters from the light curves (such as planet-to-star radius ratio $R_p/R_s$) of set 5, we followed a Markov Chain Monte Carlo (MCMC) Bayesian approach to get the distribution of probabilities of the fitted parameters and adopted the median of the distribution as the best fitted value (see \citealp{Berta2012} and Appendix for details).

To fit the white light curve, we assumed a circular orbit (eccentricity $e=0$), 
fixed the orbital period to $P = 0.81347459$ days (taken from \citealp{Blecic2013})
and set as free parameters the planet-to-star radius ratio $R_p/R_s$; the quadratic
limb darkening coefficients $u_1$, $u_2$; the central time of the transit $T_c$; the orbital 
semi-major axis over stellar radius $a/R_s$; and the orbital inclination $i$. 

In order to take into account the systematic effects induced by atmospheric extinction (and other effects observed in the 
curves obtained when we integrated the spectrum in wavelength), a third-degree polynomial with a time dependency was fitted 
to the data. Also, a first-degree polynomial with a dependency on the measured full width at half maximum (FWHM) across the 
spatial direction of each spectrum was fitted to take into account the flux variability produced by changes in seeing during 
the observations. Although this last effect is not so evident in the white light curve of set 5, it affected more strongly 
the light curves obtained using narrower filters. To be self consistent with the analysis of curves presented here, we fitted 
this seeing dependency to all the obtained light curves.

Thus, the observed flux ratio was modeled by:

\begin{equation}
 F_{transit} = \mathcal{T}_{model}(V_T) \mathcal{P}(t) \mathcal{Q}(s)
\end{equation}
where $\mathcal{T}_{model}(V_T)$ is a synthetic transit model dependent on the transit parameters $V_T$, 
$\mathcal{P}(t) = a_0 + a_1 t + a_2 t^2 + a_3 t^3$ polynomial dependent on time ($t$), and $\mathcal{Q}(s) = 1 + c_0 s$ 
a polynomial with a seeing ($s$) dependency (measured in FWHM units).

As a test, we computed the Bayesian Information Criterion (BIC, see Eq. \ref{Eq:BIC}) for the white light curve to compare the use of a second and a third-degree time dependent polynomial (plus a linear function with a seeing dependency) to fit the systematic effects present in set 5. The BIC value for the third-degree polynomial was 74 while for the second-degree polynomial it was 70, meaning that the simpler model is slightly preferred, although the fit for the curve was better using the third-degree polynomial. As a final test, we fitted the 16 light curves produced using filters of 25 nm of width using both parametrizations, the results were the same within the 1-$\sigma$ level of confidence. Since the use of a second or third-degree time dependent polynomial did not affect our results, we decided to use the later parametrization to fit all the light curves because it delivered a better fit to the systematic effects based on the RMS of the residuals.

The procedure to fit the light curve was split in a two step process. Since out of the transit we have 
$\mathcal{T}_{model}(V_T) = 1$, we first performed an MCMC simulation to obtain the coefficients of the third-degree 
time-dependent polynomial and the first degree seeing-dependent polynomial (and their corresponding 1-$\sigma$ error) 
using only the out-of-transit points. We used these values as a prior information to penalize the likelihood in order to 
fit the transit light curve model and the polynomials simultaneously in the next step of the fitting process.

The second step in order to fit the white light curve and the functions to take into account the systematic effects,
started with a $\chi^2$ minimization. To generate the transit models we used a fast Python implementation of 
the \citet{Gimenez2006} code written by Hannu Parviainen\footnote{https://github.com/hpparvi/PyTransit}, 
which has been used in previous planetary studies (e.g., \citealp{Murgas2012}, \citealp{Parviainen2013}). 
As a starting point of the minimization we used the transit parameters of \citet{Gillon2012}, the polynomial 
coefficients found in the first step, and for the limb darkening, the bolometric quadratic limb darkening 
coefficients of \citet{Claret2000}. For the limb darkening coefficients we used as a starting point of the fit 
the values of \citet{Claret2000} computed for a star with a turbulent velocity of $VT=0$ km/s, $\log g=4.5$ cm/s$^2$, 
$T_{eff}=4500$ K and $log (M/H) = 0.0$ dex; close to the values reported for WASP-43 by \citet{Hellier2011} and 
\citet{Gillon2012}. We adopted the standard deviation of normalized residuals (SDNR), which is the RMS of the points outside the transit after normalizing this baseline to unity, as the photometric error in the light curve. 
In the case of the white light curve of set 5, the SDNR of the points outside the transit were 543.7 ppm.

Once we had the best set of parameters that minimized the $\chi^2$ function, we used them as a 
starting point of the MCMC chains that computed the posterior probability distribution of the fitted the parameters. 
The total number of fitted parameters were twelve: six transit parameters, five polynomial coefficients and one 
scalar multiplying the RMS of the out-of-transit points to find out the best error of the light curve (see Appendix). 
All twelve parameters were fitted simultaneously.

To run the MCMC chains, we chose the Python code \textit{emcee} (\citealp{ForemanMackey2013}). This code 
provides a relatively fast, robust way of producing several independent chains. The different chains are 
created by slightly changing the solution that minimize the $\chi^2$ function; these different solutions will 
act as seeds to start the chains, that way we obtain a more robust result if all the chains (with different 
starting points) converge to the same probability distribution. 

We used 300 independent chains (each one with twelve free parameters) and computed 3200 iterations per chain. 
In terms of iterations, this equals to computing one chain $9.6\times 10^5$ times. To obtain the probability distribution 
from the chains, we left out the first 500 iterations as \textit{burn-in} period (the stage were the chains are still converging) 
and computed the cross-correlation length for each of the parameters. This way we only use independent samples from the chains 
to obtain the distributions. Then, each of the 300 chains with the selected independent samples, were merged into one 
in order to get the probability distribution of the fitted parameters. The final values and errors for the parameters were 
obtained by computing the percentiles of the distributions to get the median and standard deviation for each parameter. 

For the filters of several widths used to produce different light curves, the fitting process was similar 
to the one used to fit the white light curve, with the difference that we fixed two of the geometric 
transit parameters ($a/Rs$, $i$) using the results of the white light curve fit. For the broad band filters, 
we used the \citet{Claret2000} broad band quadratic limb darkening coefficients for the corresponding filters 
as a starting point for the minimization. For the smaller filters (10, 18 and 25 nm of width), we also used 
the \citet{Claret2000} broad band quadratic limb darkening coefficients as a starting point to fit the light 
curve, changing the theoretical coefficient values with the corresponding wavelength of the center of the filters.

\begin{figure}
   \centering
   \includegraphics[width=\hsize]{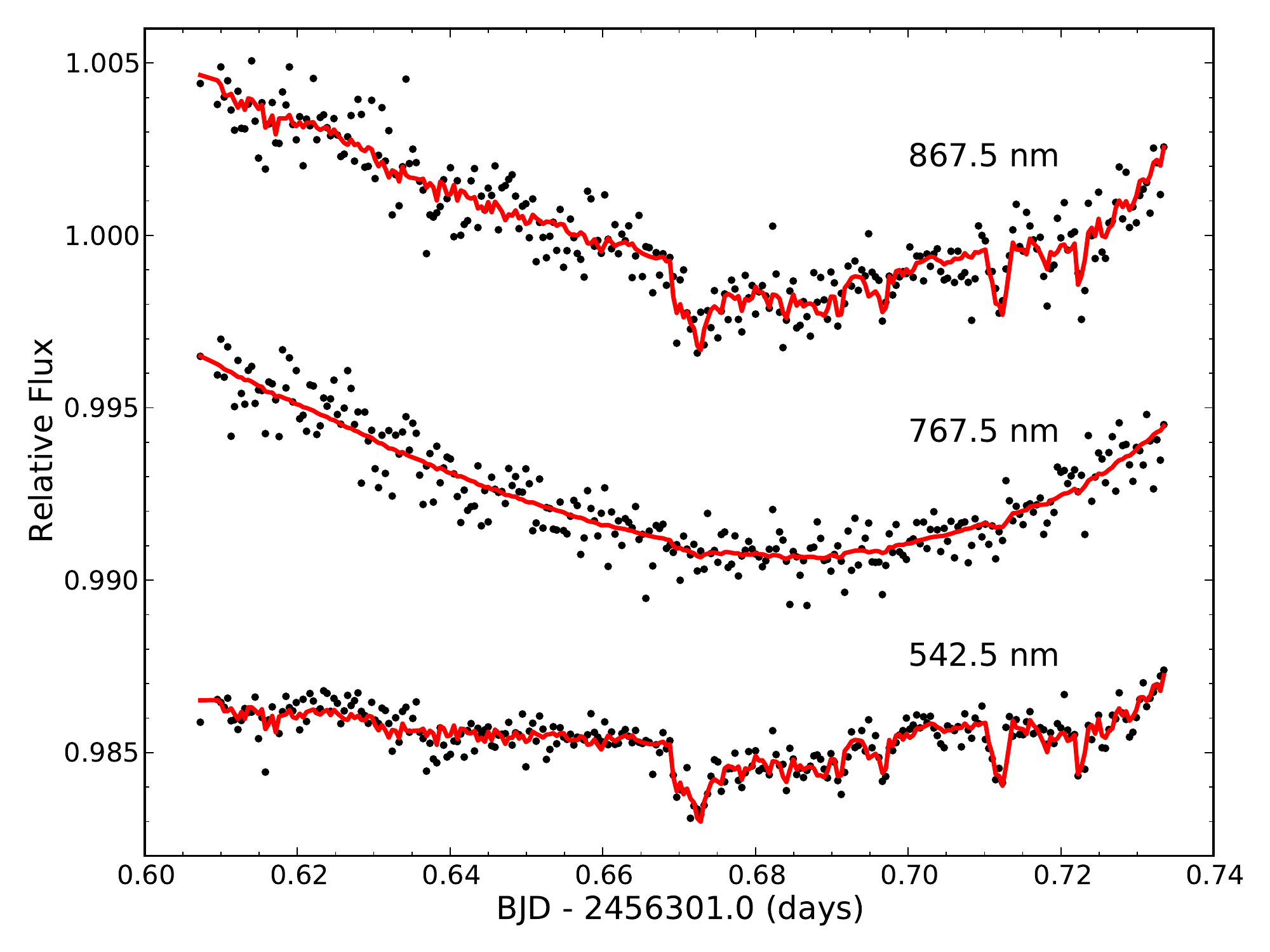}
      \caption{Example of three light curves produced using the filters of 25 nm of width for GTC set 5 with the transit removed. 
      Overplotted with a red line the best fit to take into account the systematic effects. 
      The light curves at the red and blue end of the spectrum were more affected by flux losses caused by seeing variations.}
      \label{FigPoylFits}
   \end{figure}

Figure \ref{FigPoylFits} presents three curves with the transit removed from set 5 made with filters of 25 nm of 
width and centered at 542.5, 767.5 and 867.5 nm respectively. The red line shows the best fit for the time and seeing 
dependent polynomial found by the MCMC analysis. The seeing variations affected more the curves produced using the blue 
and red wavelength regions (see Appendix). The strong improvement in the model fits when introducing the seeing dependence is indicative 
that even with a 12 arcsec slit width slit losses can still occur.

\subsection{Red noise estimation}
Time correlated noise, or red noise, can cause an underestimation of the errors in the fitted parameters of a 
light curve. To test for red noise in our data, we followed the procedure of \citet{Winn2008} which compares the 
predicted RMS of the residuals of the fitted light curve versus the measured RMS of the residuals 
of the curve binned in time. If the curve does not present red noise, the measured RMS of the residuals 
when binning the light curve in time should follow the equation:
\begin{equation}
 \sigma_N = \frac{\sigma_1}{\sqrt{N}} \sqrt{\frac{M}{M-1}}
\end{equation}
where $\sigma_1$ is the RMS of the residuals of the unbinned light curve, $\sigma_N$ is the RMS of the residuals 
of the curve binned in time, $N$ is the number of points in the binned light curve and $M$ is the number of bins. 
We measured the RMS of the residuals using $M=89$ different bin sizes between 5 and 80 minutes with a step of 
0.85 minutes and compared the measured RMS versus the predicted $\sigma_N$. For each of the light curves we computed 
$\beta = \sigma_{measured}/\sigma_N$ and adopted the median $\beta$ value for the bins 
in a time scale under 20 minutes as a multiplying factor to the errors found by the MCMC. In the cases where 
$\beta < 1$, we set $\beta \equiv 1 $ meaning that the error in those light curves was set by the MCMC procedure. 
For the white light curve we derived a $\beta_{WL}=1.5$.
   
  \begin{table}
    \caption{MCMC results of the white light transit curve of WASP-43b. The best values 
    were obtained using the posterior probability distribution found by the MCMC and the 1$-\sigma$ uncertainties 
    were computed using the MCMC values multiplied by a factor $\beta_{WL} = 1.5$ to take into account the red noise.}
    \label{table:1}      
    \centering                          
    \begin{tabular}{c c}        
    \hline\hline                 
    Parameter & Value \\    
    \hline                
    
    $R_p/R_s$ & $0.15988^{+0.00133}_{-0.00145}$ \\
    $u_1$ & $0.394 \pm 0.087$ \\
    $u_2$ & $0.289 \pm 0.119$ \\
    $T_c-2450000$ [days] & $6301.66868 \pm 5.84 \times 10^{-5}$  \\
    $i$ [deg] & $81.723 \pm 0.32$ \\
    $a/R_s$ & $4.752 \pm 0.066$ \\
    $b=(a/R_s)\cos(i)$ [$R_s$] & $0.684 \pm 0.018$ \\
    
    \hline                                   
    \end{tabular}
  \end{table}

A test performed to the residuals of the fitted light curves showed that we were able to adjust the main sources of noise present in the data (see Appendix). Hence, the use of this $\beta$ factor to increase our uncertainties should be enough to compensate the possible underestimation of errors caused by red noise.

\section{Results}
\label{sec:results}

\subsection{White light curves and timing analysis}
\label{subsec:WL}
The white light curve and the best fitted model from the MCMC procedure are presented in the top panel of 
Fig. \ref{FigWLCurve} and the obtained transit parameters are given in Table \ref{table:1}. To account for the red noise, 
the uncertainties presented in Table \ref{table:1} correspond to the 1-$\sigma$ errors from the MCMC distributions 
multiplied by the factor $\beta$ discussed in \S \ref{sec:meths} ($\beta_{WL} = 1.5$). The SDNR of the fit presented 
in the middle panel of Fig. \ref{FigWLCurve} is 372.2 ppm. The integrated flux of the target was $\approx 1.7 \times 10^8$ photons, with this number the photon noise was $79$ ppm. Comparing the photon noise to the SDNR of the whitle light curve, we can conclude that the noise level in our curve is a factor $\sim 4.7$ bigger than the theoretical limit. Using 
the parameters found for the white light curve of set 5, we derived a transit duration of $74.7\pm2.0$ minutes.

Since this planet is expected to present some degree of orbital decay caused by tidal interactions between the star and 
the planet (\citealp{Hellier2011}), it is important to determine and monitor the orbital period with high accuracy and over 
several years to detect possible variations in this parameter. With that goal in mind, we use the five white light curves 
presented in this work to obtain the central time of transit. For the fitting we use the Transit Analysis Package (TAP, 
\citealp{Gazak2012}) which implements an MCMC method to fit \citet{Mandel2002} transit models and also includes a wavelet 
method (\citealp{Carter2009}) to estimate the level of red noise in the photometric curves. 

We fitted a circular orbit to all five GTC transits, and fixed $i$, $a/R_s$, and the quadratic limb darkening coefficients 
($u_1$ and $u_2$) to the values obtained from the fit of set 5 white light curve. The transit parameter $a/R_s$ is not expected 
to change significantly during the time interval of the observations. For each epoch we allow the central time of the transit 
to vary and the results, together with 1-$\sigma$ errors, are presented in Table \ref{table:5}. The uncertainties for the central 
time of set 5 found by TAP are compatible at the 1-$\sigma$ level to the value found by our MCMC fitting process multiplied by 
the computed $\beta$ factor.

\begin{figure}
   \centering
   \includegraphics[width=\hsize]{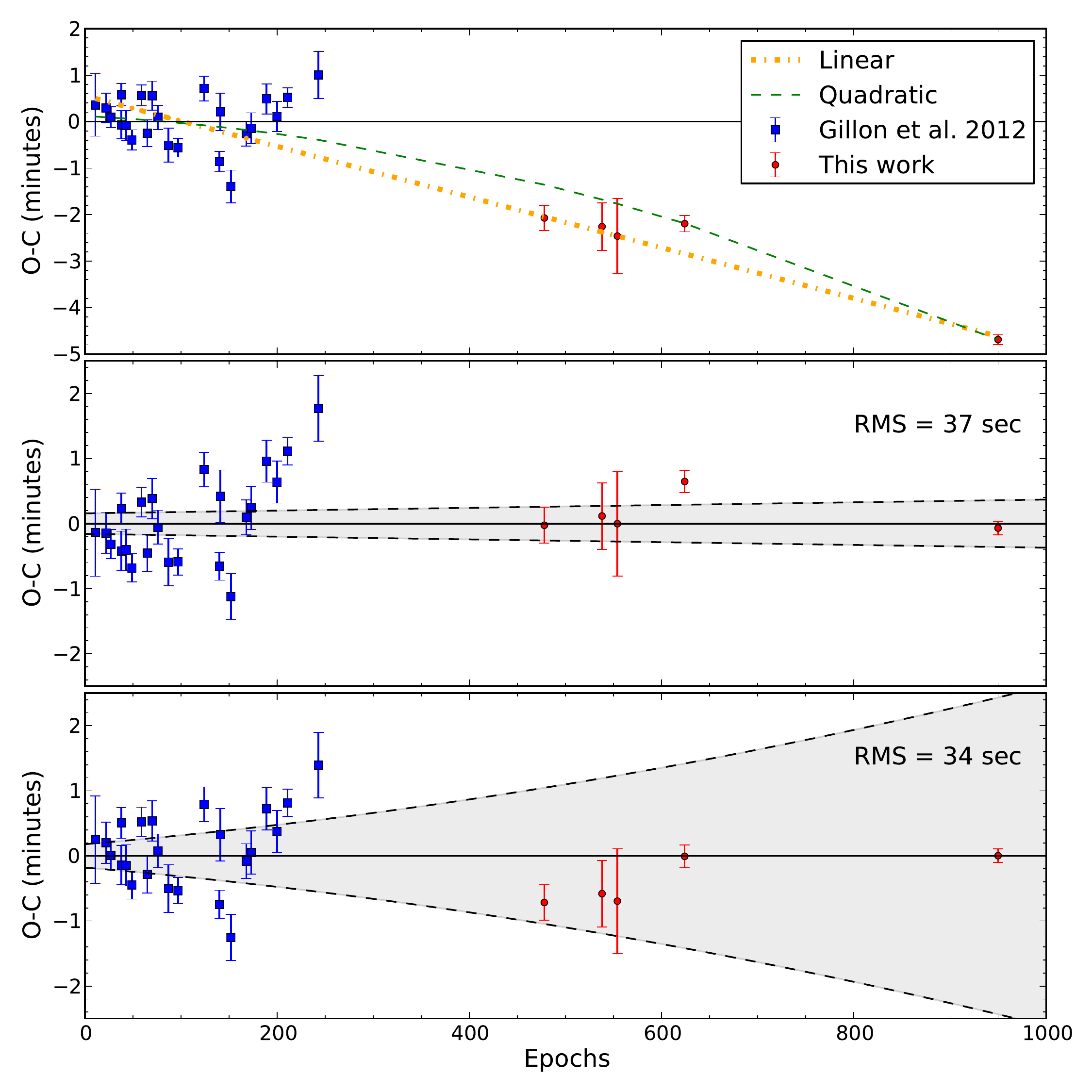}
      \caption{\textit{Top panel}: O-C diagram and linear and quadratic fits for GTC, \citet{Gillon2012} transits of WASP-43b. 
        \textit{Middle panel}: residuals of the linear fit to the data and 
      1-$\sigma$ uncertainties of fit (gray area). \textit{Bottom panel}: residuals of the quadratic fit to the 
      data and 1-$\sigma$ uncertainties of fit (gray area).}
         \label{FigTcOC}
   \end{figure}

We compute the Observed minus Calculated (O-C) diagram (Figure \ref{FigTcOC}, top panel) of the central times of the 
transits of WASP-43b using the ephemeris equation and the 23 transit midtimes presented by \citet{Gillon2012} 
together with the $T_c$ we obtained from the five GTC light curves. A clear deviation from the predicted times can 
be seen for the GTC epochs. This can be explained by the cumulative error in the ephemeris equation, specially in 
the value of the orbital period. To correct this, we fit a linear and a quadratic regression to the time residuals. Both linear and quadratic fits are weighted by the uncertainties in the central times of the transits.

After correcting for the linear trend, the updated linear ephemeris equation is:

\begin{equation}
 \label{Eq:linearPer}
  T_c = T_0 + N P
\end{equation}
where $T_0=2455528.8686144 \pm 1.1 \times 10^{-4}$ days and the resulting orbital period is 
$P=0.81347385 \pm 1.5 \times 10^{-7}$ days. The corrected O-C diagram is presented in the middle panel of Fig. \ref{FigTcOC}. 
The RMS of the timing residuals using the updated ephemeris equation is 37 seconds and $\chi^{2}_{red}$ of the linear 
fit is 5.1.

For the quadratic fit we use the same expression as \citet{Blecic2013}:
\begin{equation}
 T_c = T_0 + N P + \delta P \frac{N(N-1)}{2}
\end{equation}
where $T_0$ is a reference time, $P$ the orbital period of the planet, $N$ the number of transits since $T_0$, and 
$\delta P = \dot{P} P$ with $\dot{P}$ a term describing the variation of the period. 

The timing residual RMS is 34 seconds if we use the quadratic regression fit (bottom panel, Fig. \ref{FigTcOC}) with a 
$\chi^{2}_{red} = 4.01$. The orbital period is $P=0.81347688 \pm 8.6 \times 10^{-7}$ days, the reference time is 
$T_0=2455528.8683108 \pm 1.2 \times 10^{-4}$ days, and $\dot{P} = -0.15 \pm 0.06$ sec/year. \citet{Blecic2013} also 
found an indication of orbital decay with a $\dot{P}$ of $-0.65 \pm 0.12$ sec/year, a factor 4.3 times bigger than 
the one found in our analysis.

To compare the results of the linear and quadratic fit, we also computed the Bayesian Information Criterion (BIC):
\begin{equation}
BIC = \chi^2 + k \ln N_P
\label{Eq:BIC}
\end{equation}
where $k$ is the number of free parameters of the fit and $N_P$ is the number of data points, which favors the quadratic 
fit (BIC = 110.4) over the linear fit (BIC = 139.3). The preference of a quadratic fit to the timing data was also found by \citet{Blecic2013}.

   \begin{table}
    \caption{Transit Analysis Package (TAP) fitted central time of transit for all the GTC/OSIRIS data sets reported here.}
    \label{table:5}      
    \centering                          
    \begin{tabular}{c c}        
    \hline\hline                 
     GTC data set & $T_c-2450000.0$ (days)\\    
    \hline                        
    1 & $5917.70910 \pm 19 \times 10^{-5}$  \\
    2 & $5966.51763 \pm 36 \times 10^{-5}$  \\
    3 & $5979.53313 \pm 56 \times 10^{-5}$  \\
    4 & $6036.47675 \pm 12 \times 10^{-5}$  \\
    5 & $6301.66873 \pm 7.3  \times 10^{-5}$  \\
    \hline                                   
    \end{tabular}
  \end{table}  
   
Considering the work of \citet{Blecic2013}, this is the second study that hints to a possible orbital decay in this particular planetary system. The period and orbital decay reported in \citeauthor{Blecic2013} is a result based on the $T_c$ from \citet{Gillon2012}, their secondary transits from Spitzer, and amateur observations. We attribute the difference with our results to the different treatment of uncertainties in the reported central times due to the method used to estimate the red noise and our longer baseline of timing observations. While \citeauthor{Gillon2012} and \citeauthor{Blecic2013} used time averaging methods, we have used the wavelet method which in principle delivers more conservative estimations of uncertainties (see \citealp{Carter2009}, section 4).

We tested this difference by fitting simultaneously three complete transits of WASP-43b reported by \citet{Gillon2012} 
using TAP. We found that the uncertainties we obtain are 1.5-2.0 times larger compared with the values originally reported. 
Similar results have been obtained when comparing the wavelet with time averaging methods (\citealp{Shoyer2012}; \citealp{Shoyer2013}).

The confirmation of orbital decay could help to constrain the stellar dissipation factor $Q'_\star$ (e.g. \citealp{Rasio1996}, \citealp{Matsumura2010}) and establish the survival time of the planet. Such confirmation for WASP-43b needs to be done by a dedicated study of timing variations with a longer time baseline and a more uniform data analysis, hence the detection of orbital decay in this system remains an open question.


\subsection{Color signature}

   \begin{figure}
   \centering
   \includegraphics[width=\hsize]{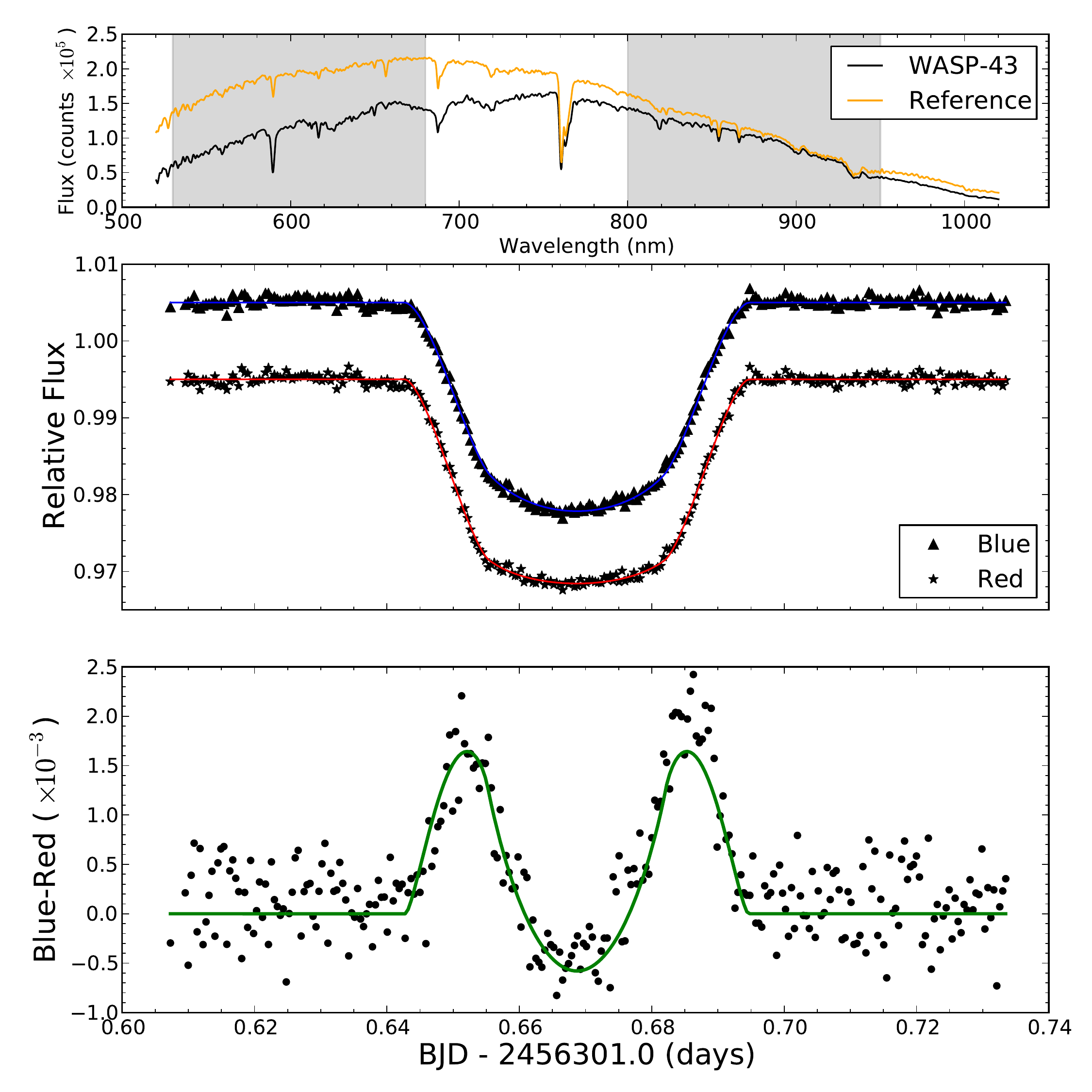}
      \caption{Color signature of WASP-43. \textit{Top panel}: Spectrum of WASP-43 and reference star showing 
       the regions were the blue and red filters were defined (gray area). 
      \textit{Middle panel}: Blue and Red light curves after correcting for systematic effects and best model fit. 
      \textit{Bottom panel}: color (Blue-Red) showing the color signature of WASP-43, the green line represents the 
      difference of the fitted models for the Blue and Red filters.}
         \label{FigColSign}
   \end{figure}
   
Simultaneous multicolor photometry of planetary transits has some very interesting applications such as the 
identification of false positives in searches for exoplanets and limb darkening studies of the host star 
(see \citealp{Tingley2004}, \citealp{Tingley2006}, \citealp{Colon2012b}, and references therein). 

As a planet transits the stellar disk it occults part of the stellar light and, due to the limb darkening effect, it will 
block more red light during the ingress and egress than during the middle of the transit. With simultaneous photometry in a 
blue and a red band it is possible to measure this effect by using the color Blue - Red measured during the transit.

\begin{figure}
   \centering
   \includegraphics[width=\hsize]{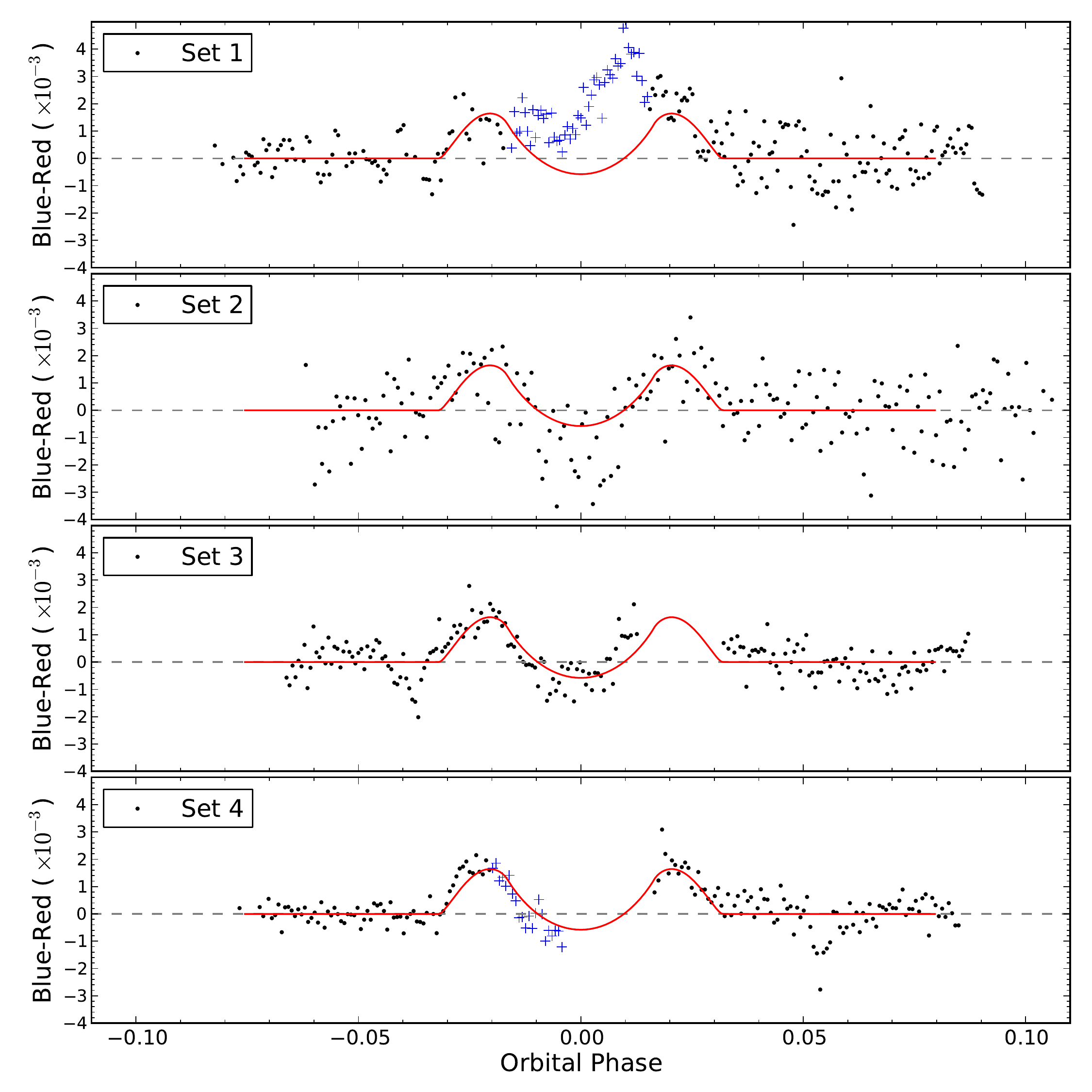}
      \caption{Color signature of WASP-43: the same filters as Fig. \ref{FigColSign} top panel but using the 
      old GTC data sets and overplotting the best fit for set 5 in red line. Blue crosses in sets 1 and 4 
      show data which were taken when the telescope presented technical issues.}
         \label{FigColSignAll}
   \end{figure}

As can be seen from in the bottom panel of Fig. \ref{FigColSign}, the effect of the color measurement (Blue-Red) of the 
transit creates horn-like structures in the ingress and egress of the transit. This effect is small, in this specific case 
with a peak near 1800 ppm. We also produced the Blue and Red light curves using the same broad band filters 
for the data sets 1 to 4, however since those observations were not complete we did not fit 
a transit model. As shown in Fig. \ref{FigColSignAll}, we detect the same horn-like signature in several of the old GTC data sets. 
The overplotted model was taken from data set 5 and is consistent with the previous GTC observations.

The fitted limb darkening coefficients for the Blue and Red curves are: $u_{1\; Blue} = 0.6057 \pm 0.057$, 
$u_{2\; Blue} = 0.1493 \pm 0.071$, $u_{1\; Red} = 0.3128 \pm 0.056$, $u_{2\; Red} = 0.2415 \pm 0.066$. The coefficients of 
the Blue curve are close to the \citet{Claret2000} values for the R Johnson filter $u_{1\; R} = 0.6029$ and 
$u_{2\; R} = 0.1418$, while the coefficients for the Red curve are slightly different to the values of \citet{Claret2004} 
for the \textit{z} sloan filter: $u_{1\; z} = 0.3834$ and $u_{2\; z} = 0.227$.

\subsection{Transmission spectroscopy}
\begin{figure}
   \centering
   \includegraphics[width=\hsize]{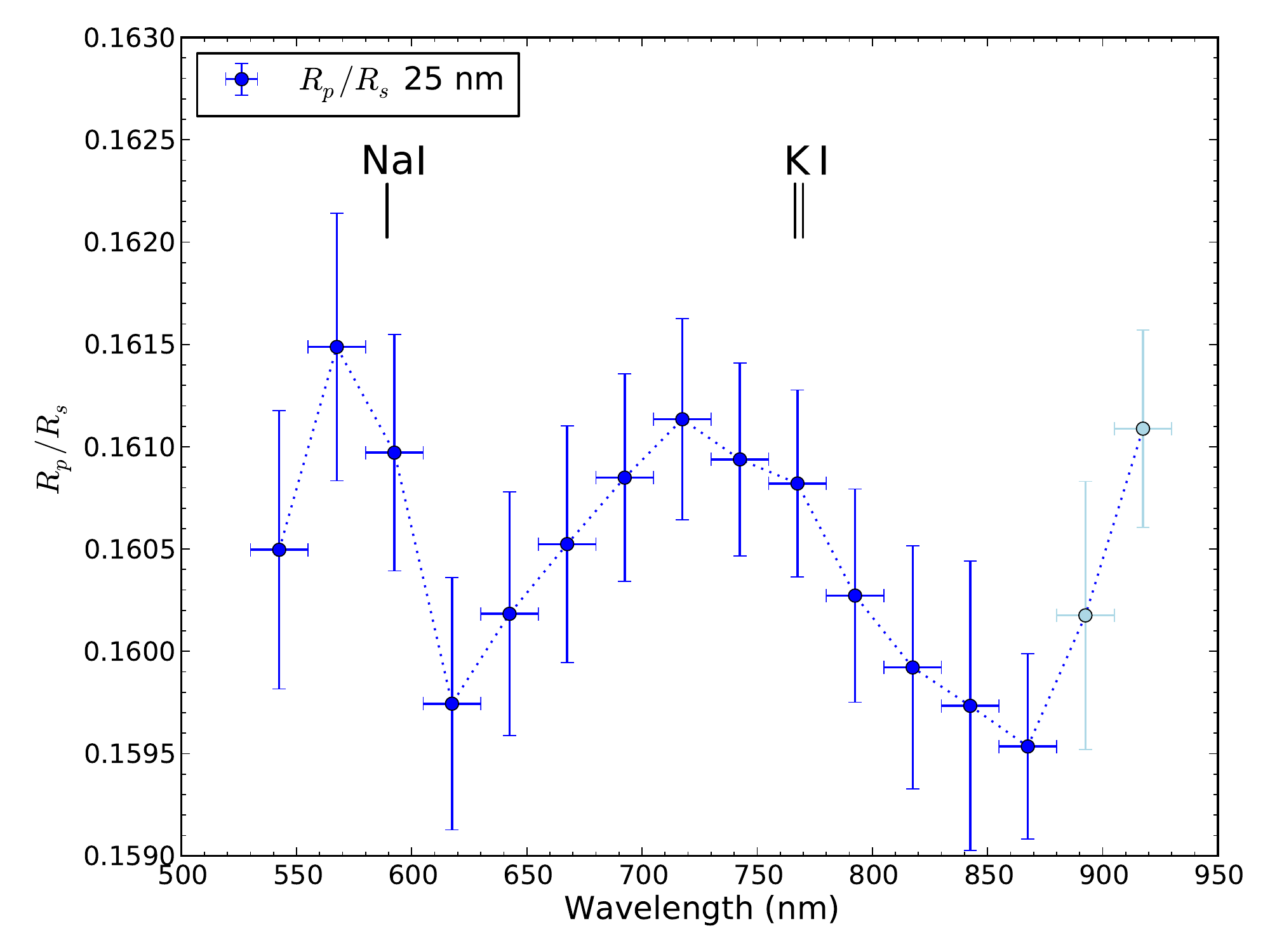}
      \caption{Transmission spectrum of WASP-43b; the horizontal bars represent the wavelength width of the filters, the vertical 
      error bar show the 1-$\sigma$ uncertainties found by the MCMC procedure after multiplying by the $\beta$ factors to take 
      into account the red noise. The last two points shown in light blue are probably affected by fringing.}
         \label{FigTransSpec}
   \end{figure}

Figure \ref{FigTransSpec} presents the results of the analysis using the filters of 25 nm of width for the 
wavelength range of 530-900 nm. Table \ref{table:2} presents the results of the measured $R_p/R_s$ for each light curve, the 
corresponding 1-$\sigma$ uncertainties found by the MCMC and the $\beta$ factors to take into account the 
red noise for each curve.

  \begin{table}
    \caption{$R_p/R_s$, 1-$\sigma$ uncertainties from the MCMC analysis and $\beta$ factors 
    for the light curves produced using the filters of 25 nm of wavelength width.}
    \label{table:2}      
    \centering                          
    \begin{tabular}{c c c}        
    \hline\hline                 
    Filter center (nm) & $R_p/R_s$ & $\beta$ \\    
    \hline     
    
    542.5 & $0.16050 \pm 0.00068$ & 1.40 \\
    567.5 & $0.16149 \pm 0.00065$ & 1.39 \\
    592.5 & $0.16097 \pm 0.00058$ & 1.24 \\
    617.5 & $0.15974 \pm 0.00062$ & 1.36 \\
    642.5 & $0.16018 \pm 0.00060$ & 1.36 \\
    667.5 & $0.16052 \pm 0.00058$ & 1.35 \\
    692.5 & $0.16085 \pm 0.00051$ & 1.20 \\
    717.5 & $0.16113 \pm 0.00049$ & 1.20 \\
    742.5 & $0.16094 \pm 0.00047$ & 1.15 \\
    767.5 & $0.16082 \pm 0.00046$ & 1.02 \\
    792.5 & $0.16027 \pm 0.00052$ & 1.21 \\
    817.5 & $0.15992 \pm 0.00059$ & 1.39 \\
    842.5 & $0.15973 \pm 0.00071$ & 1.53 \\
    867.5 & $0.15954 \pm 0.00045$ & 1.02 \\
    892.5 & $0.16018 \pm 0.00066$ & 1.46 \\
    917.5 & $0.16109 \pm 0.00048$ & 1.04 \\
    
    \hline                                   
    \end{tabular}
\tablefoot{The uncertainties are multiplyied by the factor $\beta$ presented here.}
  \end{table}

The first four bins present a maximum peak in the filter centered at 567.5 nm. 
Between 600 nm and 870 nm, the transmission spectrum presents a trend of an increase in the measured ratio $R_p/R_s$ 
towards the red part of the spectrum (between 600 nm and near 720 nm) and then the depth of the transit decreases 
until 880 nm. These trends observed with the filters of 25 nm of width were also found in a test made with 
wider filters of 75 nm of width. We speculate that the increase in $R_p/R_s$ between 620 and 720 nm 
and posterior decrease redward of 720 nm may be consistent with a planet temperature in the range 1500-2000 K and the 
presence of molecular absorption due to VO and TiO, as shown by the isothermal model computations of \citet{Fortney2010}, 
which assumed chemical equilibrium. This temperature range agrees with the derived values by \citet{Blecic2013} and 
\citet{Wang2013}. According to the model spectra of \citet{Fortney2010}, the presence of oxides make it harder to isolate 
K\,{\sc i} doublet with data of low spectral resolution; this is consistent with our finding. Detection of TiO and ViO 
at optical wavelengths has also been claimed for HD 209458b (\citealp{Desert2008}), which has a warm day side brightness 
temperature of 1320\,$\pm$\,80 K (\citealp{Crossfield2012}).

The last two bins in the redder part of the spectrum centered at 892.5 and 917.5 nm, present a higher $R_p/R_s$ when 
compared to the adjacent filters; these bins could be more affected than the rest of the spectrum by fringing, although fringing patterns are not visible in our data. According to GTC staff \footnote{http://www.gtc.iac.es/instruments/osiris/osiris.php}, fringing affects only at a 5\% level for wavelengths $\lambda > 930$ nm and less than 1\% for wavelengths $\lambda < 900$ nm, but 
given the high precisions that transmission spectroscopy requires, it could be an important source of noise for red
wavelengths when using the GTC R1000R grism.

To test the results obtained with filters of 25 nm of width and to search for possible atmospheric features that the use 
of broad filter may have overlooked, we proceeded to create light curves using narrower filters in the regions near the Na\,{\sc i} 
doublet, the K\,{\sc i} doublet and the region between these two features were some telluric signals are located.
   
   \begin{figure*}
   \centering
   \includegraphics[width=\hsize]{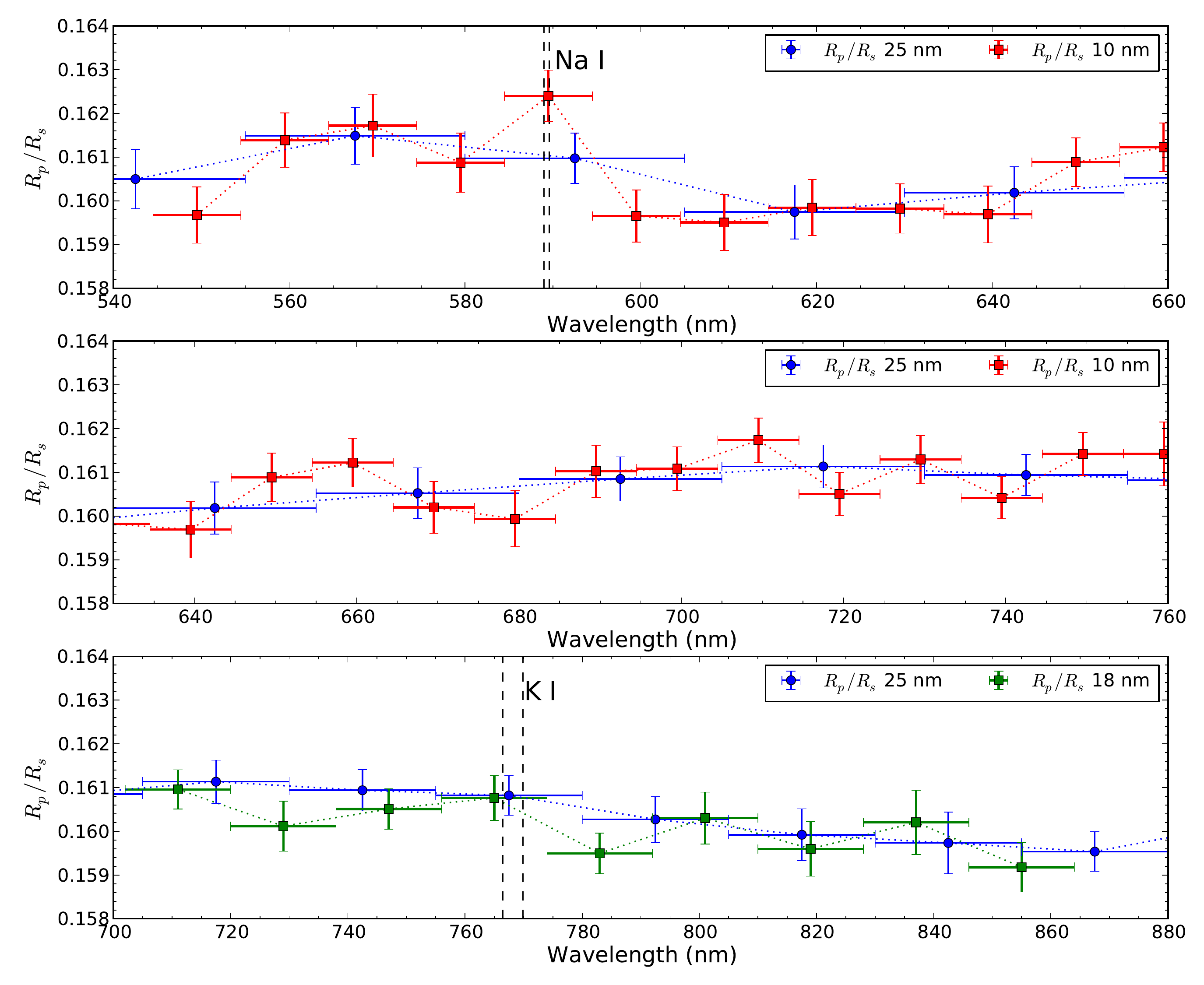}
      \caption{Transmission spectrum of WASP-43b, the horizontal bars represent the width of the filters used
      to bin the spectrum and the vertical error bars represent the errors found by the MCMC procedure and 
      the $\beta$ factors to take into account the red noise. The gray line shows the extracted spectrum of WASP-43 
      in an arbitrary scale. \textit{Top panel}: zoom to the Na\,{\sc i} doublet region showing the 25 nm (blue points) 
      and 10 nm width filters results (red points); \textit{Middle panel}: same for the region between the Na and 
      K lines; \textit{Bottom panel}: K\,{\sc i} doublet region showing the 25 nm and 18 nm filters results.}
         \label{FigSodium}
   \end{figure*}

Figure \ref{FigSodium}, top panel, presents the comparison between the results using the light curves obtained 
with the 25 nm and 10 nm width filters for the region near the Na\,{\sc i} doublet (see Table \ref{table:3}). 
We can see that the 10 nm width filter that possess the same center as the Na{\sc i} doublet hints to an excess in 
the measured $R_p/R_s$ when compared to the adjacent redder filters, however the filters in the blue 
region next to the same line present a similar level of planetary radius than the one found in the Na feature. 
Using only the filter centered at 599.5 nm next to the Na{\sc i} doublet, the difference between 
these two filters is $\Delta (R_p/R_s)_{Na-599.5} = (2.75 \pm 0.84)\times 10^{-3}$; if we use the filter centered at 579.5 nm 
we find a less significant excess of $\Delta (R_p/R_s)_{Na-579.5} = (1.53 \pm 0.90)\times 10^{-3}$. If we compute the weighted average $(R_p/R_s)$ using the bins centered at 579.5 and 599.5 nm, we have $(R_p/R_s)_{Avg} = 0.16022 \pm 4.51\times 10^{-4}$. Using this average value, $\Delta (R_p/R_s)_{Na-Avg} = (2.18 \pm 0.74)\times 10^{-3}$, a 2.9-$\sigma$ of significance excess. Assuming a stellar radius of 
$R_s = 0.667 R_\odot$ (\citealp{Gillon2012}), this excess in Na could be traced back to an atmospheric height of 
$\Delta (R_p)_{Na-Avg} = 1012 \pm 344$ Km. All these tests points to an excess in the measured planet-to-star radius ratio that 
could be produced by the atmosphere of WASP-43b, but our 2.9-$\sigma$ precision is not enough to claim an unambiguous detection. This tentative detection could be improved with new observations using, for example, 
GTC/OSIRIS R1000B grism which covers a bluer part of the spectrum. Sodium has been detected in other exoplanets with slightly colder 
temperatures than WASP-43b like HD 209458b (\citealp{Charbonneau2002}), XO-2b (\citealp{Sing2012}) with a temperature of 
1500 K (\citealp{Machalek2009}), and HD 189733b (\citealp{Redfield2008}) with an atmospheric temperature of $1340 \pm 150$ K 
(\citealp{Lecavelier2008}).
   \begin{table}
    \caption{$R_p/R_s$, 1-$\sigma$ uncertainties from the MCMC analysis and $\beta$ factors 
    for the light curves produced using the filters of 10 nm of wavelength width near the Na\,{\sc i} 588.9 and 589.5 nm doublet.}
    \label{table:3}      
    \centering                          
    \begin{tabular}{c c c}        
    \hline\hline                 
    Filter center (nm) & $R_p/R_s$ & $\beta$\\    
    \hline                        
    
    549.5 & $0.15967 \pm 0.00064$ & 1.09 \\
    559.5 & $0.16138 \pm 0.00062$ & 1.14 \\
    569.5 & $0.16172 \pm 0.00072$ & 1.27 \\
    579.5 & $0.16087 \pm 0.00068$ & 1.10 \\
    589.5 & $0.16240 \pm 0.00059$ & 1.03 \\
    599.5 & $0.15965 \pm 0.00060$ & 1.10 \\
    609.5 & $0.15951 \pm 0.00064$ & 1.22 \\
    619.5 & $0.15984 \pm 0.00064$ & 1.17 \\
    629.5 & $0.15982 \pm 0.00056$ & 1.00 \\
    639.5 & $0.15969 \pm 0.00065$ & 1.23 \\
    649.5 & $0.16088 \pm 0.00055$ & 1.12 \\
    
    \hline                                   
    \end{tabular}
\tablefoot{The uncertainties are multiplyied by the factor $\beta$ presented here.}
  \end{table}    
The central panel of Fig. \ref{FigSodium} shows an intermediate region between the Na\,{\sc i} and K\,{\sc i} doublet where 
we produced light curves using filters of 10 nm of width. The planet-to-star radius ratio of the small filters are in 
agreement with the results found using filters of 25 nm of width and present hints of an excess in $R_p/R_s$ at 
the filters centered at 658 nm (near H$_\alpha$) and 708 nm, but due to the errors in $R_p/R_s$ we can not claim a detection of 
the planet atmosphere in this region.
   
  \begin{table}
    \caption{$R_p/R_s$, 1-$\sigma$ uncertainties from the MCMC analysis and $\beta$ factors 
    for the light curves produced using the filters of 18 nm of wavelength width near the K\,{\sc i} 766.5 and 769.9 nm 
    doublet.}
    \label{table:4}      
    \centering                          
    \begin{tabular}{c c c}        
    \hline\hline                 
    Filter center (nm) & $R_p/R_s$ & $\beta$ \\    
    \hline                        
    
    711.0 & $0.16096 \pm 0.00045$ & 1.02 \\
    729.0 & $0.16012 \pm 0.00057$ & 1.29 \\
    747.0 & $0.16051 \pm 0.00046$ & 1.03 \\
    765.0 & $0.16076 \pm 0.00051$ & 1.00 \\
    783.0 & $0.15950 \pm 0.00046$ & 1.04 \\
    801.0 & $0.16030 \pm 0.00059$ & 1.25 \\
    819.0 & $0.15960 \pm 0.00062$ & 1.35 \\
    837.0 & $0.16021 \pm 0.00074$ & 1.47 \\
    855.0 & $0.15918 \pm 0.00057$ & 1.14 \\
    
    \hline                                   
    \end{tabular}
\tablefoot{The uncertainties are multiplyied by the factor $\beta$ presented here.}
  \end{table}

  \begin{figure*}
   \centering
   \includegraphics[width=\hsize]{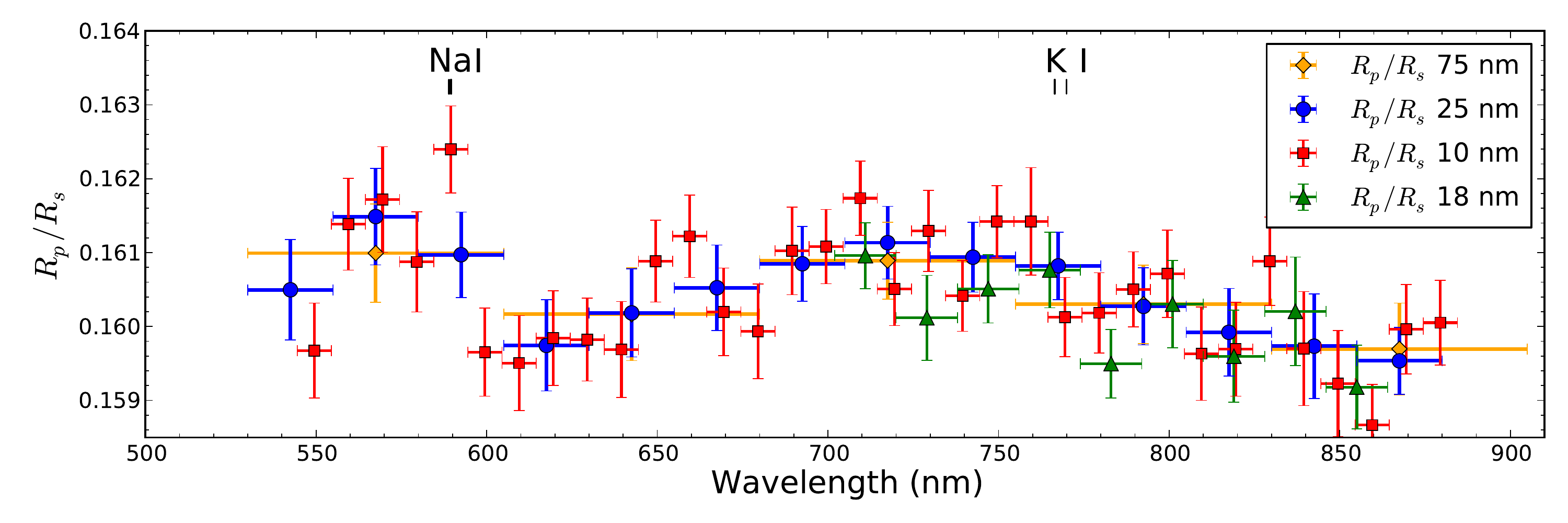}
      \caption{Transmission spectrum of WASP-43b showing all the wavelength bins used in this study. The horizontal bars 
      represent the wavelength width of the filters, the vertical error bar show the 1-$\sigma$ uncertainties found by 
      the MCMC procedure after multiplying by the $\beta$ factors to take into account the red noise.}
      \label{FigAllBins}
  \end{figure*}

The bottom panel of Fig. \ref{FigSodium} presents the wavelength region near K\,{\sc i} doublet. The telluric absorption feature caused by $O_2$ (see Fig. \ref{FigExtrSpec}) is blended with the 766.5 and 769.9 nm Potassium doublet at the resolution of our spectrum. One needs to be very careful not to confuse telluric variability from 
possible planetary signature (see Appendix). Thus, for this wavelength region, we chose a wider than the 
usual 10 nm filter used in the regions with shorter wavelengths in order to cover the entire oxygen and potassium lines. Taking the weighted average of the planet-to-star radius ratio of the two 18 nm of width filters adjacent to the K\,{\sc i} doublet 
(747.0 nm and 783.0 nm) and using those bins as a pseudo-continuum, we found a difference between the line and the continuum of 
$\Delta (R_p/R_s)_{K-Avg} = (0.755 \pm 0.61)\times 10^{-3}$, meaning that we do not detect a statistical significant excess in 
the measured $R_p/R_s$ in the filter centered at the Potassium.

As a summary, Fig. \ref{FigAllBins} presents all the wavelength bins used to create the different light curves of 
WASP-43b. With our MCMC analysis and red noise estimation, the measured planet-to-star radius ratio agrees well for light curves 
created using filters of different width in wavelength.


\section{Conclusions and final remarks}
\label{sec:conclusions}

In this work, we presented the results of GTC/OSIRIS long-slit spectroscopic observations of the extra-solar planet 
WASP-43b. Four partial transits taken in December 22 2011, February 8 2012, February 21 2012, and April 18 2012; and 
one full transit event observed in January 8 2013 are analyzed. The observed wavelength range was 510-1040 nm and, in 
the case of the January 8 2013 set, with a resolution of 374-841 at 751 nm.

Integrating the full range of wavelength available in the observed spectrum of the target and 
its reference star, we produced a white light photometric curve. We fitted a transit model to the data set 5 (full transit 
observations) and, using an MCMC analysis, we obtained transit parameters with their corresponding uncertainties that take 
into account the level of white and red noise in the photometric curve.

For each of the five GTC/OSIRIS WASP-43b data sets we fit the central time of the observed individual transits. Combining 
these results with previous observations made by \citet{Gillon2012}, we see a decrease in time in 
the O-C versus epoch diagram. We attributed this trend to a cumulative error in the period determination ephemeris and 
proceeded to fit a linear and quadratic function to determine a new period. Using a linear fit, we find a new orbital period 
of $P=0.81347385 \pm 1.5 \times 10^{-7}$ days. As reported by \citet{Blecic2013}, the O-C data are better fitted using a quadratic function, with this parametrization we find an orbital period of $P=0.81347688 \pm 8.6 \times 10^{-7}$ days and a change in the period of $\dot{P} = -0.15 \pm 0.06$ sec/year. The $\dot{P}$ that we reported here is smaller than the one found by \citet{Blecic2013}. Based on this discrepancy, a dedicated study of the transit timing of this system made over several years needs to be carried out in order to confirm the orbital decay of this particular planet.

Using Blue and Red light curves, we detected the classical horn-like structures representative of the color signature of WASP-43. 
We confirmed the color signature detection using the old GTC data sets. This shows the potential of performing multicolor 
studies using long slit spectroscopy and transiting planets.

We also present here the transmission spectrum of WASP-43b using filters of 25, 18, and 10 nm of width. With the wider filters, 
we observed a possible excess in $R_p/R_s$ near the position of the Na\,{\sc i} doublet, a trend of increasing planet-to-star 
radius ratio from blue to red wavelengths between 600 and 720 nm, and a decreasing $R_p/R_s$ between 720 and 880 nm. The 
increase in $R_p/R_s$ values between 620 and 720 nm and the decrease in $R_p/R_s$ redward of 720 nm may be consistent with the 
temperature of WASP-43b and the presence of VO and TiO predicted by atmospheric models.
 
For the region near Na\,{\sc i} 588.9 and 589.5 nm doublet, we found an excess in $R_p/R_s$ using filters of 10 nm of width. 
Taking the average of the $R_p/R_s$ of the two adjacent filters as a continuum, we obtained a difference of 
$\Delta (R_p/R_s)_{Na-Avg} = (2.18 \pm 0.74)\times 10^{-3}$; meaning that the excess is detected with 2.9-$\sigma$ of confidence. 
The regions in the blue part of the Na\,{\sc i} doublet were more affected by flux losses due to seeing variations. In the region between 640 and 760 nm, we observed an increase in the measured planet-to-star radius ratio until the spectrum reach the Potassium doublet. 

We probed the wavelength region near the K\,{\sc i} doublet (766.5 nm, 769.9 nm) using filters of 18 nm of width. We do not 
detect a significant excess in the planet-to-star radius ratio at the position of the Potassium doublet. After 
the K\,{\sc i} doublet and until 855 nm, the bins follow a trend of a decreasing planet-to-star radius ratio.

According to the models of \citet{Fortney2010} Sodium and Potassium features are strong at planetary temperatures between 
1000 and 1500 K, close to the brightness temperature found by \citet{Blecic2013} of $T=1684 \pm 24$ K and $T=1485 \pm 24$ K for 
the 3.6 and 4.5 $\mu$m Spitzer bands. Our data points to an excess in the measured planet-to-star radius in the Na\,{\sc i} 
doublet and no detection of an excess in the K\,{\sc i} doublet, which would make WASP-43b a planet with atmospheric 
characteristics similar to HD 209458b.


\begin{acknowledgements}
      Based on observations made with the Gran Telescopio Canarias (GTC), installed in the Spanish 
      Observatorio del Roque de los Muchachos of the Instituto de Astrof\'isica de Canarias, 
      in the island of La Palma. S.H. acknowledges financial support from the Spanish Ministry of 
      Economy and Competitiveness (MINECO) under the 2011 Severo Ochoa Program MINECO SEV-2011-0187.
      All the figures presented here were made using Matplotlib (\citealp{Hunter2007}).
\end{acknowledgements}


\bibliographystyle{aa}
\bibliography{biblio.bib}


\begin{appendix} 
\label{sec:appendix}
\section{A note about the extraction of the spectrum, data modeling, and red noise}
\subsection{Data reduction}
During the data analysis of this work we found that the $R_p/R_s$ measured in the filters centered at the Potassium doublet were 
sensitive to the removal of bad pixels and/or cosmic rays. The IRAF routine used to extract the spectrum, APALL, presents the 
option to remove and replace deviant pixels. The user can choose the threshold level to replace the pixels that present a lower 
sensitivity (bad pixels) or high values caused by cosmic rays. Since the O$_2$ telluric feature is fairly deep, a lower 
threshold limit wrongfully identified the pixel that was near the minimum flux of the absorption line as a deviant pixel and 
replaced that pixel value. This occurred more or less randomly for both target and reference star in the time series due to the changes 
in the depth of the telluric line caused by atmospheric variability. This effect produced extra noise in the light curve computed 
using a wavelength width of 25 and 18 nm around the K doublet. 

The extra noise produced by the cleaning algorithm created a deeper and noisier transit in that filter and, with our MCMC, a spurious 
detection of Potassium in WASP-43b. This problem was fixed by reducing the data several times fine tuning the threshold level of 
rejection in order to detect and remove bad pixels and cosmic rays but not replacing the values of the deepest points in the O$_2$ 
absorption line. This delivered a higher quality light curve centered near the Oxygen telluric line that presented a similar noise 
levels seen in the curves of the adjacent wavelength regions.

   \begin{figure}
   \centering
   \includegraphics[width=\hsize]{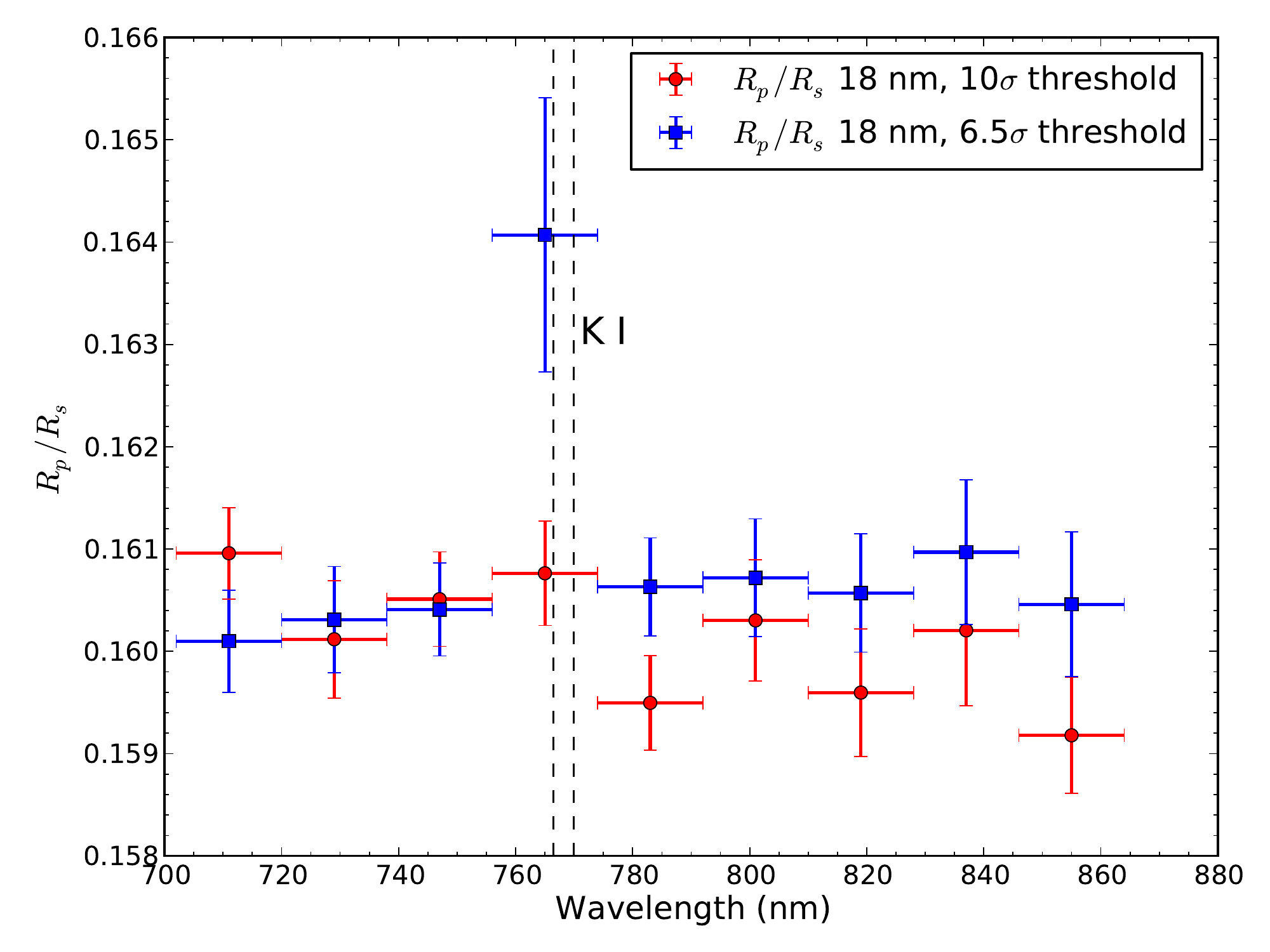}
      \caption{Comparison of results of the measured $R_p/R_s$ around the Potassium doublet using different threshold levels to 
      remove bad pixels and cosmic rays, and no seeing correction in the light curve fitting process. A lower threshold level 
      could produced a spurious detection of an excess in the planet-to-star radius ratio near a deep telluric line as shown 
      by blue squares. Both blue and red data points do not present the trend in decreasing $R_p/R_s$ between 750 and 870 nm as 
      opposed of the results shown in Fig. \ref{FigTransSpec} and \ref{FigSodium}.}
         \label{FigKComparison}
   \end{figure}
   
It is also important to correctly parametrize the model to correct any flux variation that is correlated with some of the 
observational parameters like the seeing. In our first attempt to fit the data we did not consider seeing as a source of noise 
in the measured flux ratio between the target and reference star; that caused that the MCMC computed very large error bars in 
the measured $R_p/R_s$ in the filters computed in the blue part of the spectrum and a flat transmission spectrum between 750 
and 870 nm.
   
Figure \ref{FigKComparison} shows a comparison of the results obtained using two different threshold levels to remove bad pixels 
and cosmic rays: a 6.5$\sigma$ shown in red and the final 10$\sigma$ rejection limit in blue. Both results were obtained without 
correcting for seeing variations and present a flat transmission spectrum in the red region next to the K doublet, as opposed of 
the results presented here in Fig. \ref{FigTransSpec} and \ref{FigSodium}.

\subsection{Data analysis}
As we explained in \S 3.3, we fitted a transit model adding two polynomials to take into account the time and seeing dependent 
systematic effects present in the data:
\begin{equation}
 F_{transit} = \mathcal{T}_{model}(V_T) \mathcal{P}(t) \mathcal{Q}(s)
\end{equation}
where $\mathcal{T}_{model}(V_T)$ is a synthetic transit model dependent on the transit parameters $V_T$. For the 
white light curve $V_T = (R_p/R_s,u_1,u_2,T_c,a/Rs,i)$ with $R_p/R_s$ the planet-to-star radius ratio, $(u_1,u_2)$ the 
quadratic limb darkening coefficients, $T_c$ the central time of the transit, $a/R_s$ the semi-major axis over stellar 
radius, and $i$ the orbital inclination. $\mathcal{P}(t)$ is a time dependent polynomial, and $\mathcal{Q}(s)$ a seeing dependent 
polynomial:
\begin{equation}
 \mathcal{P}(t) = a_0 + a_1 t + a_2 t^2 + a_3 t^3
\end{equation}
\begin{equation}
 \mathcal{Q}(s) = 1 + c_0 s
\label{Eq:seeingAppendix}
\end{equation}

For the MCMC fitting procedure we followed a similar approach as \citet{Berta2012} and used as  likelihood $\mathcal{L}$
\begin{equation}
 \ln \mathcal{L} = -N \ln( p ) - \frac{\chi^2}{2p^2}
\end{equation}
where $N$ is the number of points in the curve, $p$ a coefficient to normalize the $\chi^2$. The function $\chi^2$ 
compares the data points with the model :
\begin{equation}
 \chi^2 = \sum_{i=1}^{N} \left( \frac{d_i-m_i}{\sigma_i} \right)^2
\end{equation}
where $d_i$ is the data point, $m_i$ the model point, and $\sigma_i$ the error in the measurement which in our case was assumed 
to be the SDNR of the points outside the transit. 

The probability priors used for each parameter are presented in Table \ref{table:ap1}. The use of a normal prior 
for the polynomial parameters to take into account the systematic effects was adopted using the information of the MCMC analysis 
performed using the points outside the transit.
  \begin{table}
    \caption{Type of probability priors used in the analysis.}
    \label{table:ap1}      
    \centering                          
    \begin{tabular}{c c}        
    \hline\hline                 
     Parameter & Prior \\    
    \hline                        
 
    $R_p/R_s$ & Jeffreys\\
    $u_1$     & Normal  \\
    $u_2$     & Normal  \\
    $T_c$     & Uniform \\
    $a/R_s$   & Jeffreys\\
    $i$       & Uniform \\
    $a_0$     & Normal  \\
    $a_1$     & Normal  \\
    $a_2$     & Normal  \\
    $a_3$     & Normal  \\
    $c_0$     & Normal  \\
    $p$       & Uniform \\
    
    \hline                                   
    \end{tabular}
  \end{table}

\begin{figure*}
   \centering
   \includegraphics[width=\hsize]{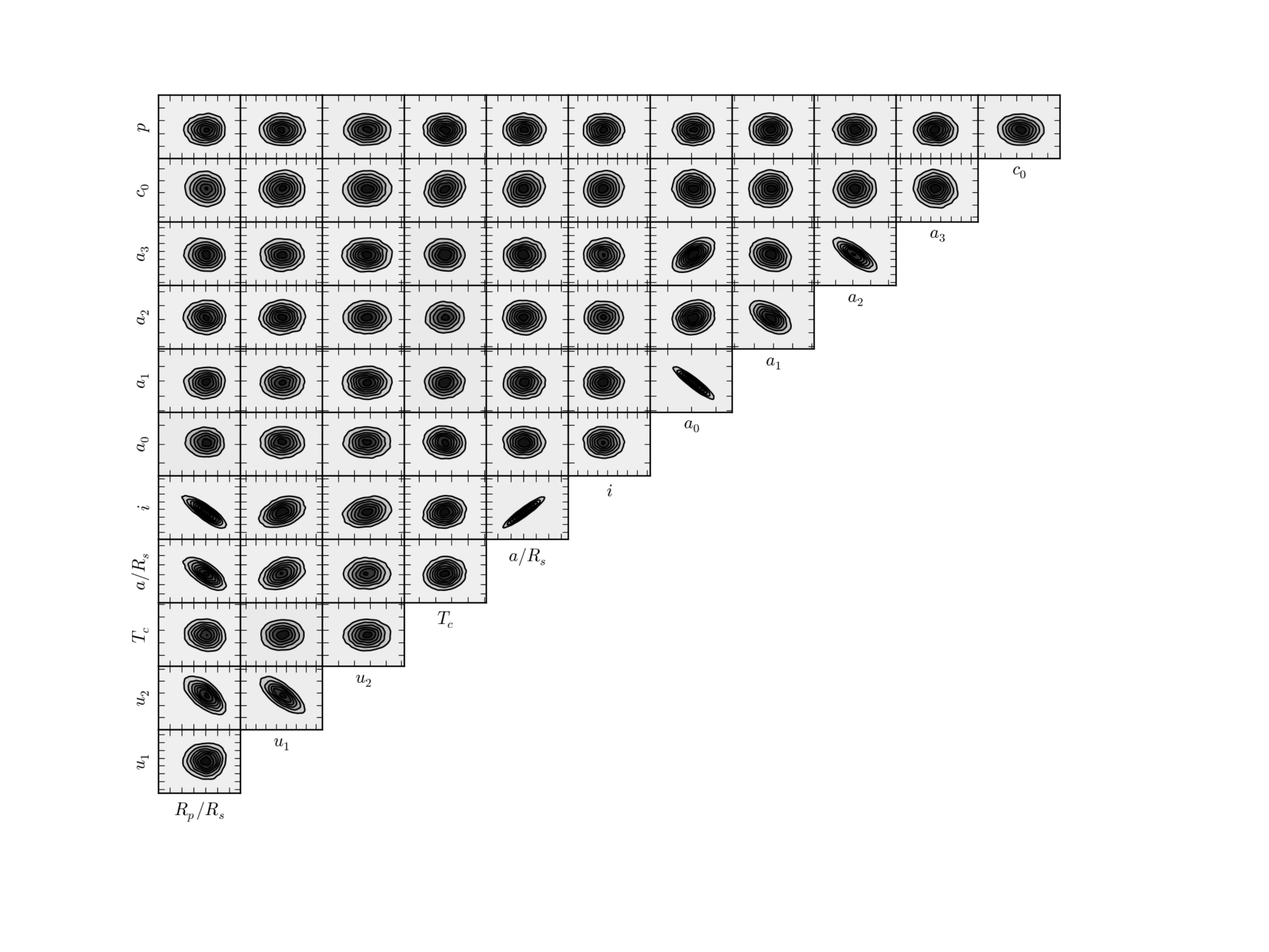}
      \caption{Correlation plot showing the posteriori distribution of the parameters fitted to the white light curve of WASP-43b.}
         \label{FigCorrelation}
   \end{figure*}

Figure \ref{FigCorrelation} presents a correlation plot for the posteriori distribution of all the parameters used to fit the white light curve. The parameters that are more correlated with the planet-to-star radius ratio are the limb darkening coefficient $u_2$, the semi-major axis over stellar radius $a/R_s$, and the orbital inclination $i$.

\begin{figure}
   \centering
   \includegraphics[width=\hsize]{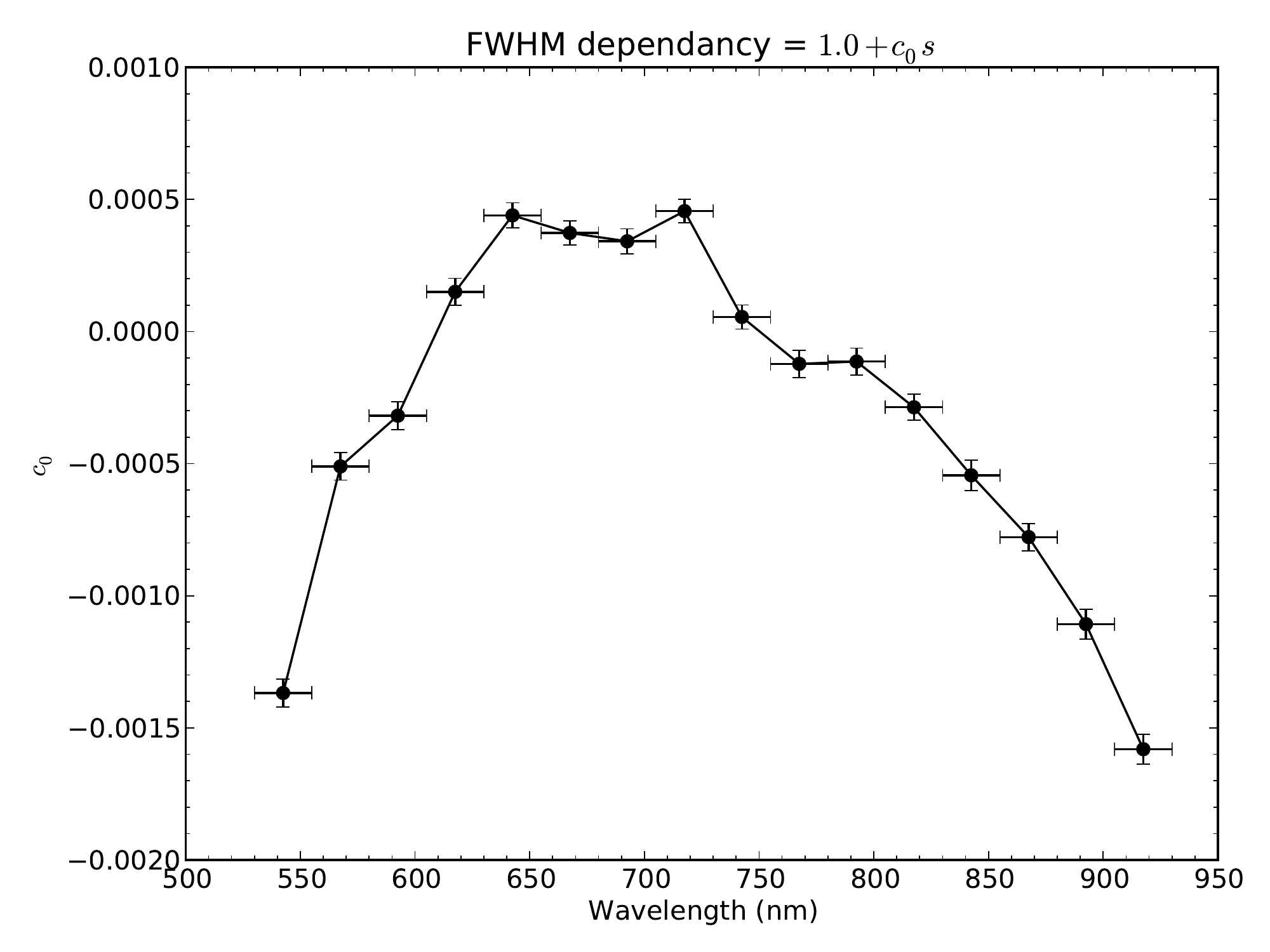}
      \caption{Seeing coefficient $c_0$ as function of wavelength. The flux losses produced by seeing variations were more strong at both blue and red ends of the detector.}
         \label{FigFwhmvsLambda}
   \end{figure}

Figure \ref{FigFwhmvsLambda} shows the variation of the coefficient $c_0$ used to fit the seeing variations according to Eq. \ref{Eq:seeingAppendix}. The coefficient varies across wavelength, with a greater dependency on FWHM at both ends of the wavelength range of the detector.

\subsection{Red noise analysis}
To compare our error estimation with TAP (\citealp{Gazak2012}), we performed an analysis of the residuals of the fitted light curves using our MCMC method. TAP is based in the paper of \citet{Carter2009} to compute the red noise contribution in the light curves. Their method is applied when the square of the Fourier transform of the residuals ($\mathcal{S}$) follow a power law:

\begin{equation}
\mathcal{S} = \frac{A}{f^\gamma}
\end{equation}
with $A$ a constant, $f$ the Fourier frequency, and $\gamma$ the exponent of the power law. If $\gamma=0$ we are in the presence of white noise, $\gamma=1.0$ pink noise, and $\gamma=2.0$ red noise.

\begin{figure}
   \centering
   \includegraphics[width=\hsize]{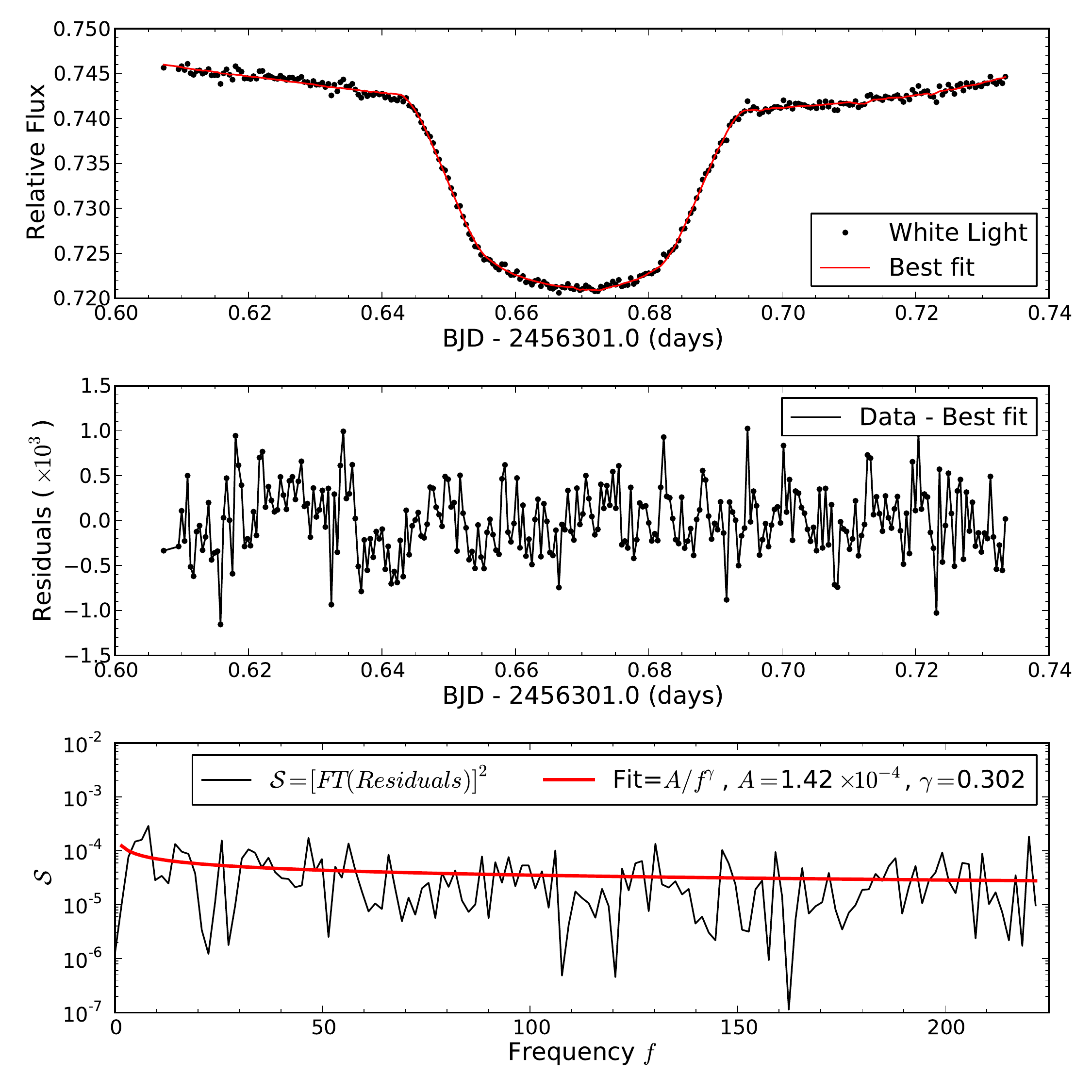}
      \caption{Red noise analysis. \textit{Top panel}: White light curve of WASP-43b and best fit. \textit{Middle panel}: Residuals of the white light curve (curve after subtracting the best fit). \textit{Bottom panel}: Square of the Fourier transform of the residuals. A fit of the form $A/f^\gamma$ yields an exponent value of $\gamma=0.302$; according to \citet{Carter2009} this means that the major noise source in this curve is white noise.}
         \label{FigFFTRedNoise}
   \end{figure}

We computed $\mathcal{S}$ for the white light curve (see Fig. \ref{FigFFTRedNoise}) and the curves produced using the filters of 75 and 25 nm of width. In all the cases the residuals followed a power law with $\gamma \lesssim 0.3$, indicating that with our fitting procedure the residuals are dominated by white noise. We can also conclude that this $\gamma \lesssim 0.3$ regime is far from the assumed noise level of TAP (with $\gamma=1.0$), meaning that in this particular case TAP could be overestimating the uncertainties. Since our fitting procedure is close to the white noise regime, we think that our estimation is more accurate than using TAP for this specific data set.

\end{appendix}


\end{document}